\documentclass[twocolumn,superscriptaddress,amsmath,amssymb,aps,prb,floatfix]{revtex4-2}

\usepackage{graphicx}% Include figure files
\usepackage{float}
\usepackage{dcolumn}% Align table columns on decimal point
\usepackage{bm}% bold math
\usepackage{hyperref}% add hypertext capabilities
\usepackage[all]{hypcap}
\usepackage{siunitx}% package for SI units
\usepackage{upgreek}% package to get upright greek letters
\usepackage{soul}
\usepackage{textgreek}

\makeatletter

\newcommand{\Rmnum}[1]{\expandafter\@slowromancap\romannumeral #1@}
\makeatother

\newcommand{\bea}{\begin{eqnarray}}
\newcommand{\eea}{\end{eqnarray}}
\newcommand{\bpm}{\begin{pmatrix}}
\newcommand{\epm}{\end{pmatrix}}
\newcommand{\bal}{\begin{aligned}}
\newcommand{\eal}{\end{aligned}}

\newcommand{\e}{\mathrm{e}}
\newcommand{\ii}{\mathrm{i}}

\sethlcolor{yellow}
\setstcolor{red}

\makeatother

\begin{document}

\title{Ring-Exchange Interaction Effects on Magnons in Dirac Magnet CoTiO$_3$}% 
\author{Yufei Li}
\affiliation{Department of Physics, The Ohio State University. Columbus, OH 43210}
\author{Thuc T. Mai}%
\affiliation{Quantum Measurement Division, Physical Measurement Laboratory, NIST, Gaithersburg, MD 20899}
\author{M. Karaki}%
\affiliation{Department of Physics, The Ohio State University. Columbus, OH 43210}
\author{E.V. Jasper}%
\affiliation{Department of Physics, The Ohio State University. Columbus, OH 43210}
\author{K.F. Garrity}%
\affiliation{Materials Measurement Science Division, Materials Measurement Laboratory, NIST, Gaithersburg, MD 20899}
\author{C. Lyon}%
\affiliation{Department of Physics, The Ohio State University. Columbus, OH 43210}
\author{D. Shaw}%
\affiliation{Department of Physics, Colorado State University, Fort Collins, CO 80523}
\author{T. DeLazzer}%
\affiliation{Department of Physics, Colorado State University, Fort Collins, CO 80523}
\author{A.J. Biacchi}
\affiliation{Nanoscale Device Characterization Division, Physical Measurement Laboratory, NIST, Gaithersburg, MD 20899}
\author{R.L. Dally}
\affiliation{NIST Center for Neutron Research, National Institute of Standards and Technology, Gaithersburg, MD 20899}
\author{D.M. Heligman}%
\affiliation{Department of Physics, The Ohio State University. Columbus, OH 43210}
\author{J. Gdanski}%
\affiliation{Department of Physics, The Ohio State University. Columbus, OH 43210}
\author{T. Adel}%
\affiliation{Quantum Measurement Division, Physical Measurement Laboratory, NIST, Gaithersburg, MD 20899}
\author{M.F. Mu\~noz}%
\affiliation{Quantum Measurement Division, Physical Measurement Laboratory, NIST, Gaithersburg, MD 20899}
\author{A. Giovannone}%
\affiliation{Department of Physics, The Ohio State University. Columbus, OH 43210}
\author{A. Pawbake}
\affiliation{Laboratoire National des Champs Magnetiques Intenses, LNCMI-EMFL. Grenoble, France}
\author{C. Faugeras}
\affiliation{Laboratoire National des Champs Magnetiques Intenses, LNCMI-EMFL. Grenoble, France}
\author{J.R. Simpson}%
\affiliation{Department of Physics, Astronomy, and Geosciences, Towson University, Towson, MD 21252}
\affiliation{Quantum Measurement Division, Physical Measurement Laboratory, NIST, Gaithersburg, MD 20899}
\author{K. Ross}%
\affiliation{Department of Physics, Colorado State University, Fort Collins, CO 80523}
\author{N. Trivedi}%
\affiliation{Department of Physics, The Ohio State University. Columbus, OH 43210}
\author{Y.M. Lu}%
\affiliation{Department of Physics, The Ohio State University. Columbus, OH 43210}
\author{A.R. Hight Walker}%
\affiliation{Quantum Measurement Division, Physical Measurement Laboratory, NIST, Gaithersburg, MD 20899}
\author{R. Vald\'es Aguilar}
\email{valdesaguilar.1@osu.edu}
\affiliation{Department of Physics, The Ohio State University. Columbus, OH 43210}

\begin{abstract}
The magnetic interactions that determine magnetic order and magnon energies typically involve only two spins. While rare, multi-spin interactions can also appear in quantum magnets and be the driving force in the ground state selection and in the nature of its excitations. By performing time-domain terahertz and magneto-Raman spectroscopy measurements combined with theoretical modeling, we determine the origin of the magnon excitation gap in Dirac antiferromagnet CoTiO$_3$. By adding a ring-exchange interaction in a hexagonal plaquette of the honeycomb lattice to both an XXZ spin model and to a low energy spin-orbital flavor wave model, a gap is generated in the magnon spectrum at the Brillouin zone center. With this addition, the flavor wave model reproduces a large swath of experimental results including terahertz, Raman, inelastic neutron scattering, and magnetization experiments.
\end{abstract}

\date{\today}% It is always \today, today,
\maketitle

In magnetically ordered materials with localized electrons, the fundamental magnetic interactions result from the exchange of electrons \cite{Heisenberg1926,Heisenberg1928,Dirac1929}. Typically, only the interaction between pairs of electrons' spins is considered to explain the nature of the ground state and its excitations, whereas three-, four-, and six-spin interactions are ignored. When these higher order interactions occur in a loop they are called cyclic or ring exchange. Such interactions have only been documented in a few cases: bulk and thin films of solid $^3$He \cite{3HeKummer1975,3HeGodfrin1980,3HeOsheroff1980,RMPRoger1983,RogerJPCS2005}; in the high-\textit{T}$_\mathbf{c}$ parent compound La$_2$CuO$_4$ \cite{SugaiRaman1990,ColdeaPRL2001}; more recently in a honeycomb cobaltite \cite{KrugerCobaltite2023}, and in a kagome magnet \cite{OrensteinRing2023}. 

Here, we provide an additional instance of the importance of ring-exchange on a quantum material. We use a combination of time domain THz spectroscopy (TDTS) and magneto-Raman spectroscopy to measure the temperature and magnetic field dependence of the low energy magnetic excitations in CoTiO$_3$. In this proposed Dirac topological magnon material \cite{yuan2020dirac,neutron2021}, the origin of the energy gap in the magnon spectrum at the Brillouin zone center has remained undetermined until now. Our study provides a detailed examination of the two, out of four, lowest energy magnons. We deduce that the gap opens due to the ring-exchange interaction between the six Co$^{2+}$ spins on a hexagon. This interaction also explains the selection rules of the THz magnon absorption and their magnetic field dependence. Finally, we clarify that topological surface magnons are not expected in CoTiO$_3$. Our results highlight the importance of the small, but finite, many-spin interactions on the magnetic ground state and its excitations.

The classification of topological materials has brought about a new perspective on the properties of solid state systems \cite{topologicalinsulatorreview,weylsemimetalreview}. This has been used to discover new electronic topological states of matter and has increased the possibilities for producing material properties by design \cite{Po2017,topologicalquantumchemistry}. More recently, the pursuit of materials where the interplay between topology and magnetism generates topological magnon excitations has begun \cite{topologicalmagnon}. CoTiO$_3$, for example, is proposed to have a topological Dirac crossing between its magnon bands. While new materials have been predicted to host topologically protected magnons on their surfaces\cite{luandlu,lu2022}, their conclusive experimental demonstration remains an open problem. Here we address whether CoTiO$_3$ can host such topologically protected surface magnons.

\begin{figure}[t]
  \centering
    \includegraphics[width=0.45\textwidth]{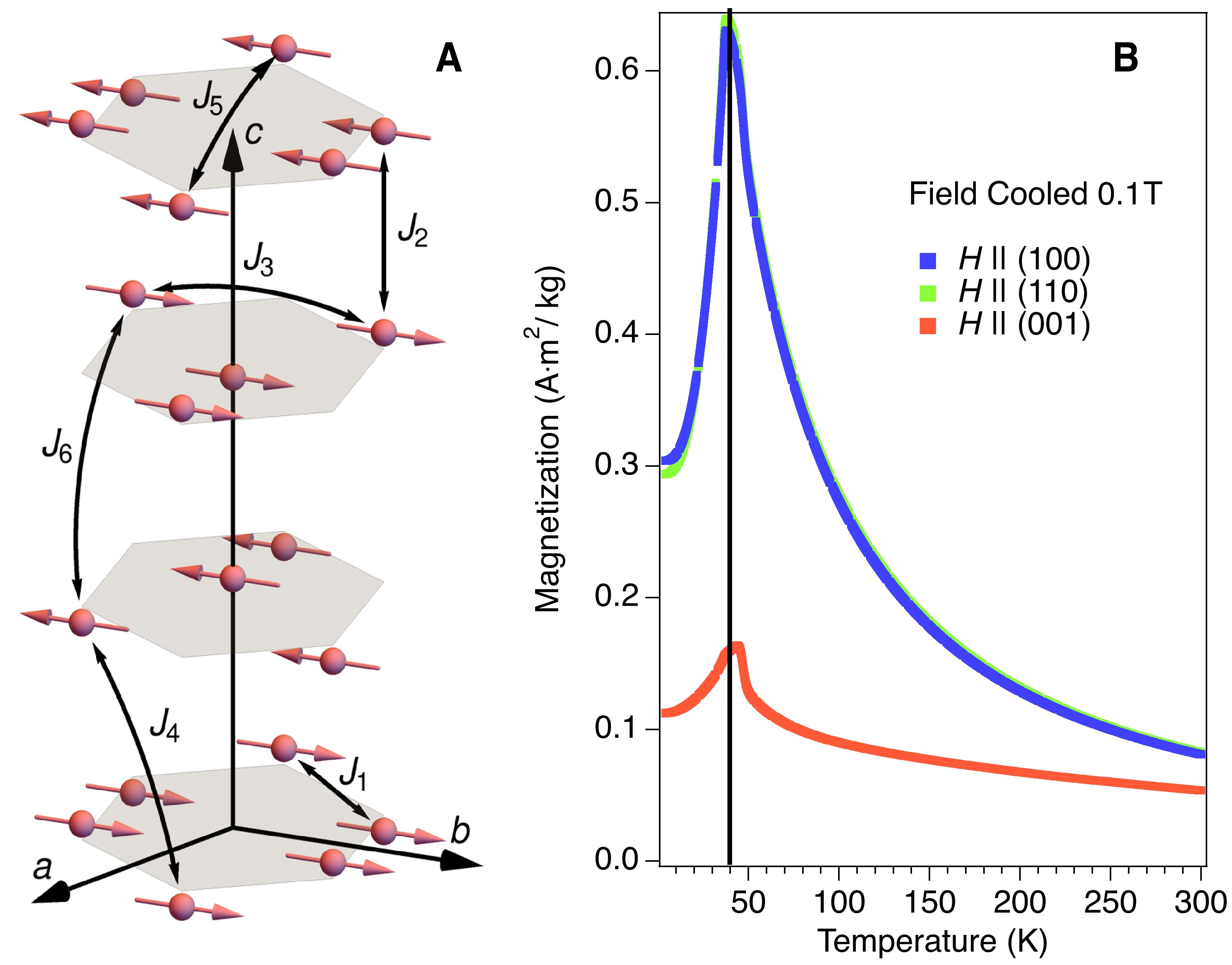}
    \caption{\textbf{Magnetic structure and magnetization in CoTiO$_3$. A.} Magnetic structure of CoTiO$_3$ as determined by elastic neutron scattering with lattice constants $a=b=$ 5.064 \r{A} and $c=$ 13.91 \r{A} \cite{yuan2020dirac,neutron1964,neutron2021}. Parallel Co spins reside on the vertices of buckled honeycomb planes stacked antiferromagnetically along the $c$-axis. Magnetic pair interactions $J_1$ through $J_6$ are indicated. \textbf{B.} Magnetization parallel (blue and green) and perpendicular (red) to the honeycomb plane of the single crystals used for the TDTS and Raman measurements. The N\'eel temperature is indicated by the black vertical line.}
    \label{Fig1}
\end{figure}

CoTiO$_3$ has a layered structure with space group $R\bar{3}$. The Co atoms form a buckled honeycomb lattice in the hexagonal plane, where each atom is either slightly above or below a plane. These ferromagnetic planes are stacked antiferromagnetically along the $c$ [001] axis \cite{Xray1958,neutron2021}, see Fig. \ref{Fig1}A, so-called type-A antiferromagnetism. An antiferromagnetic phase transition occurs at the N\'eel temperature of $T_{\mathrm{N}}=\SI{38(3)}{\kelvin}$\cite{chi1958,heatcapacity2011,permeability2017,permeability2021}. In this phase, the Co spins are ferromagnetically aligned within the honeycomb plane but, due to the lattice symmetry, the spin direction cannot be uniquely determined \cite{neutron1964,neutron2021}. This direction is typically described to be along the $b$ axis [010].

The magnetic degrees of freedom for Co$^{2+}$ ions in a trigonally distorted octahedral environment are composed of spin and orbital angular momenta where $S=3/2$ and the effective $l=1$, respectively \cite{AandBbook}. Taking into account spin-orbit coupling and the distorted octahedral environment, the local atomic electronic states form six doublets where the lowest energy state can be considered to have an effective spin $\tilde{S}=1/2$ \cite{Sano2018kitaevheishenberg,liu2018pseudospin,yuan2020dirac}. The interactions between neighboring Co$^{2+}$ have been calculated in this octahedral environment \cite{liu2018pseudospin,Sano2018kitaevheishenberg,liu2020kitaev}. When projected into the $\tilde{S}=1/2$ space, these interactions generate Heisenberg, Kitaev, and off-diagonal symmetric exchange couplings. The theoretical phase diagram for a two-dimensional honeycomb lattice of Co$^{2+}$ atoms includes a Kitaev quantum spin liquid \cite{KITAEV20062}, vortex, zigzag, and ferromagnetic phases\cite{liu2020kitaev}. CoTiO$_3$ is in the latter phase where these ferromagnetic planes stack antiferromagnetically.

\begin{figure*}[tb]
%\begin{center}%  
    \centering
    \includegraphics[width=0.85\textwidth]{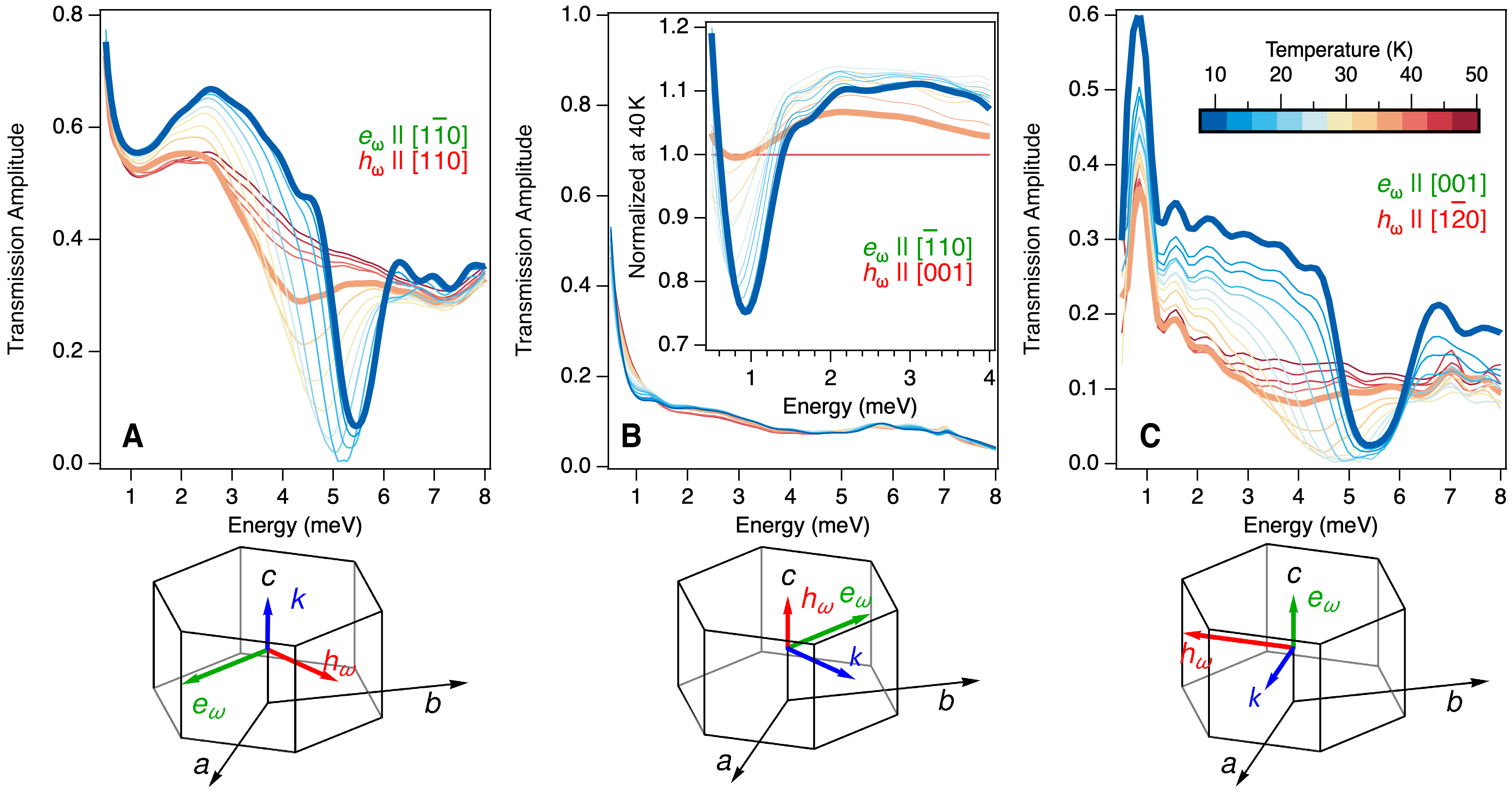}
    \caption{\textbf{THz magnon spectrum of CoTiO$_3$}. Each of the three panels is composed of the THz transmission spectrum on top and on the bottom is the polarization configuration drawn with respect to the non-Bravais hexagonal lattice indicating the direction of the THz electric ($\bm{e}_\omega$) and magnetic ($\bm{h}_\omega$) fields and their propagation direction ($\bm{k}$). \textbf{A.} Transmission spectrum for $\bm{e}_\omega$ and $\bm{h}_\omega$ both in the honeycomb plane showing one prominent absorption mode at $\approx\SI{5.4}{meV}\approx\SI{1.3}{T\hertz}$. \textbf{B.} Spectrum with $\bm{h}_\omega$ perpendicular to the honeycomb plane and $\bm{e}_\omega$ parallel to it with no strong features present. The inset shows the transmission spectra normalized to the $\SI{40}{\kelvin}$ spectrum. An absorption mode $\approx\SI{0.9}{meV}\approx\SI{0.23}{T\hertz}$ is active below $T_\mathrm{N}$. \textbf{C.} Transmission spectra with $\bm{h}_\omega$ parallel to the honeycomb plane and $\bm{e}_\omega$ perpendicular to it, these show the same mode identified in \textbf{A}. The data on panels \textbf{A} and \textbf{C} show that the $\SI{5.4}{meV}$ magnon is only active when $\bm{h}_\omega$ is parallel to the honeycomb plane, making it a magnetic dipole mode. Because the $\SI{0.9}{meV}$ is only observed in the \textbf{B} configuration, we cannot determine if it is active for other directions of $\bm{e}_\omega$ and $\bm{h}_\omega$.}
    \label{Fig2}
%\end{center}
\end{figure*}

This $\tilde{S}=1/2$ approximation in the XXZ Hamiltonian captures many of the features seen by inelastic neutron scattering in CoTiO$_3$ \cite{yuan2020dirac,neutron2021}. However, it is insufficient to explain the details of the low energy magnetic scattering spectra. In an XXZ Hamiltonian, the spin components in the honeycomb plane interact with different strength than the $z$-components. These experiments also found that first spin-orbital excited state has finite dispersion in momentum space, and that its energy is only twice as large as the bandwidth of the low energy excitations \cite{yuan2020spin,neutron2021}. In addition, the magnon spectrum of the effective $\tilde{S}=1/2$ spins in the XXZ Hamiltonian is gapless at the zone center, while experiments indicate that there is a gap of approximately $\SI{1}{\milli\electronvolt}$\cite{yuan2020dirac,neutron2021}. These observations imply that the effective $\tilde{S}=1/2$ picture of the XXZ Hamiltonian within linear spin wave theory is not sufficient to fully describe the magnetic excitations in CoTiO$_3$. While a quantum order-by-disorder mechanism \cite{Rau2018} was proposed in ref.\cite{neutron2021} to explain the opening of the gap in CoTiO$_3$, its magnitude is too small compared to the experimentally estimated one. Below, we resolve these issues with the combination of the first detailed TDTS and magneto-Raman spectroscopy study of the low energy magnetic dynamics in single crystals of CoTiO$_3$ as a function of temperature and magnetic fields, accompanied by an expanded theoretical model.

\section{Results}

Figure \ref{Fig1}\textbf{B} shows the magnetization of the single crystals on which TDTS and Raman experiments were performed (details of the crystal growth and magnetization measurements are given in ref. \cite{SI}). A clear antiferromagnetic phase transition is detected at $\SI{38}{\kelvin}$, identified by a vertical line in the figure. No temperature hysteresis was observed, indicating a continuous phase transition, as found earlier \cite{heatcapacity2011,permeability2017,permeability2021}. 

In TDTS transmission measurements, magnon absorption is identified by the decrease of the transmission in a narrow frequency range centered at the magnon frequency below the magnetic ordering transition temperature \cite{ThucSFSO,MWarrenGVS} (details of TDTS are given in ref. \cite{SI}). We describe below the magnon absorptions observed in CoTiO$_3$ and show them in figure \ref{Fig2}. Panels \textbf{A}-\textbf{C} show three different polarization configurations of the THz electric ($\bm{e}_\omega$) and magnetic ($\bm{h}_\omega$) fields for three different crystal cuts. Panels \textbf{A} and \textbf{C} show that, when $\bm{h}_\omega$ is parallel to the honeycomb plane, one main absorption mode dominates the spectra below $T_\mathrm{N}$ with an energy of approximately $\SI{5.4}{meV}\approx\SI{1.3}{T\hertz}$ at $\SI{8}{\kelvin}$. This energy coincides with one of the magnon modes at the zone center reported previously \cite{yuan2020dirac,neutron2021}, which thus confirms this magnetic dipole absorption as a magnon. The shallow minimum around $\SI{1}{meV}$ in figure \ref{Fig2}\textbf{A} is present for all temperatures shown even above $T_\mathrm{N}$, and thus it cannot be ascribed to a magnon absorption.

The inset in figure \ref{Fig2}\textbf{B} shows an additional low energy absorption that also coincides with a magnon observed in neutron spectroscopy, though we determine its energy to be $\approx\SI{0.9}{meV}\approx\SI{0.23}{T\hertz}$. This is slightly lower than estimated in ref.\cite{neutron2021} and different from a previous antiferromagnetic resonance measurement on a powder sample \cite{AFMresonance}. This mode is referred to as a pseudo-Goldstone mode and corresponds to the small oscillatory deviations of the magnetic moment in the honeycomb plane. These deviations cost no energy when there is no in-plane anisotropy. The selection rules for this mode cannot be completely determined because it could only be observed weakly for the combination of $\bm{h}_\omega\parallel$ [001] and $\bm{e}_\omega$ $\parallel$ to the honeycomb plane (figure \ref{Fig2}\textbf{B}). If the mode is purely magnetic dipole, then it should be only observable in this configuration.

\begin{figure}[!t]
  \centering
    \includegraphics[width=0.42\textwidth]{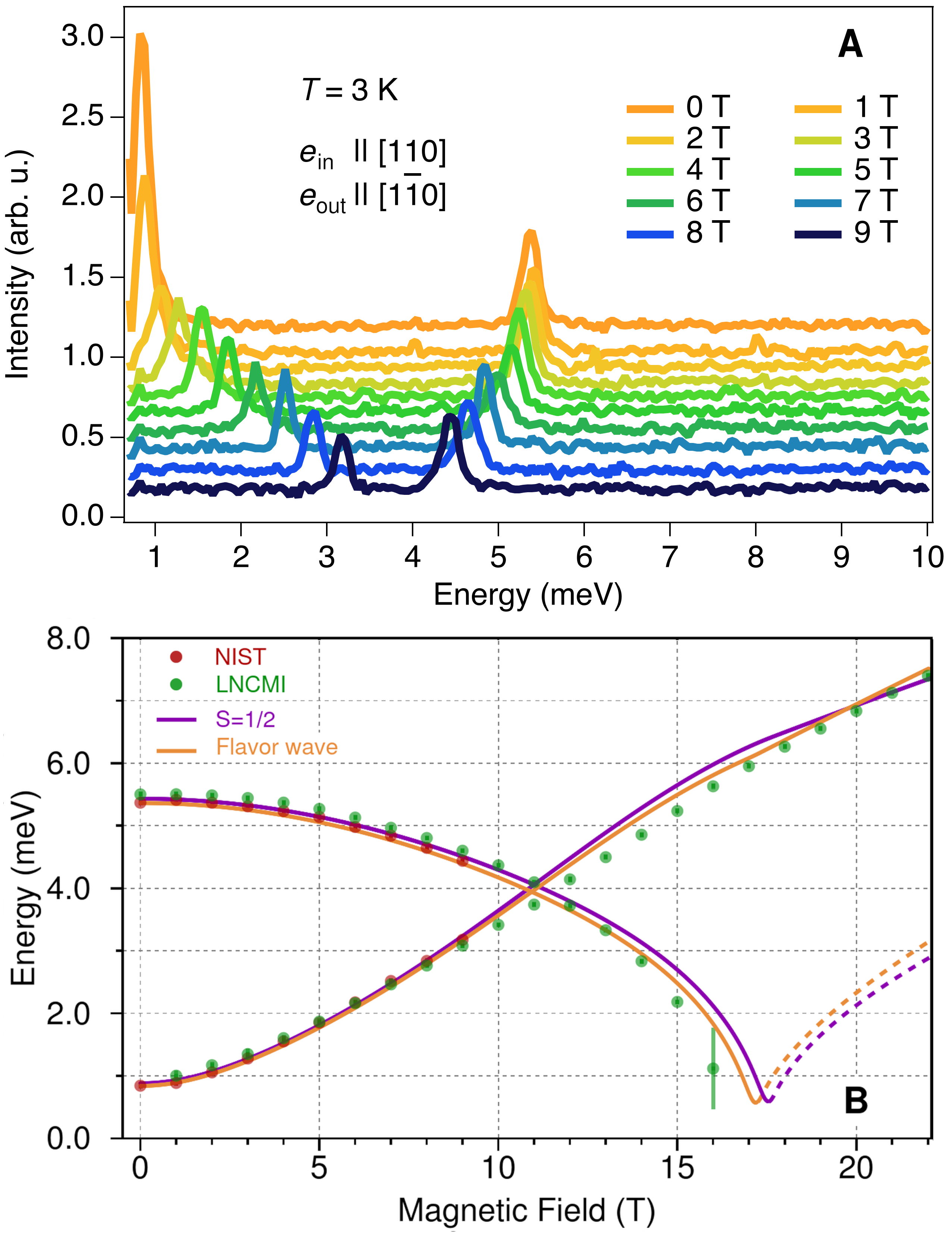}
    \caption{\textbf{Magnetic field dependence of Raman spectra.} \textbf{A.} Raman spectra at 3 K of the two lowest energy magnons for fields up to 9 T applied parallel to the honeycomb plane. \textbf{B.} Fit of the field dependence of the magnon energies using both the $\tilde{S}=1/2$ model (purple) and the flavor wave model (orange). The dashed lines above $\sim$17 T indicate that this mode has finite momentum and is unobservable with Raman scattering.}
    \label{Fig4}
\end{figure}

Raman scattering experiments show the same two magnons observed with TDTS at zero magnetic field but, due to the narrower peaks in the Raman experiment, the energies of these magnons are determined more precisely: $\SI{0.82(5)}{\milli\electronvolt}$ and $\SI{5.37(5)}{\milli\electronvolt}$, details are given in ref. \cite{SI}. Figure \ref{Fig4}\textbf{A} shows the results of the magneto-Raman scattering experiment for magnetic fields smaller than 9 T with orthogonal (crossed) polarization configuration, with $\bm{e}_{\mathrm{in}}$ and $\bm{e}_{\mathrm{out}}$ denoting incident and scattered linear polarizations. The measurement was carried out on a crystal cut perpendicular to the $c$-axis, the same crystal whose THz data are shown in fig. \ref{Fig2}\textbf{A}. The external static magnetic field is applied parallel to the $a$-$b$ plane. In addition to the magnons at low energy, several modes at higher energy become active below $T_\mathrm{N}$. These correspond to the spin-orbital excitations reported before \cite{neutron2021,yuan2020spin} and to additional phonons that become active at the zone center because of the doubling of the unit cell size with magnetic order. The detailed investigation of these modes, including their response to applied magnetic fields will be reported elsewhere \cite{thuccotio3}.

In a magnetic field applied parallel to the honeycomb plane, the Raman spectra show that the energies of the two lowest magnon modes shift significantly, as shown in figure \ref{Fig4}\textbf{A}. For larger magnetic fields up to 22 T in the hexagonal plane, a separate experiment was performed in the high magnetic field laboratory (LNCMI) in Grenoble, France with unpolarized light at a sample temperature of 5 K. Figure \ref{Fig4}\textbf{B} shows the results of fitting the low energy Raman peaks both at low and high magnetic fields with Gaussian functions to obtain the magnetic field dependence of their energies (circle markers), along with model fitting results (lines) that will be discussed in the next section. The magnon energies seem to cross around $\approx\SI{11}{\tesla}$ indicating little to no coupling between these magnons. As the field increases, their energies separate further until one magnon reaches the lowest measurable energies around $\approx\SI{17}{\tesla}$. The signal of this magnon is completely absent above this field, marking a phase transition to the fully polarized state. A magnetic field up to $\SI{7}{\tesla}$ applied along the $c$-axis does not shift the magnons in any measurable way as shown in ref. \cite{SI}.

\begin{table*}[t] 
\caption{\textbf{Models for magnon gap opening and magnetic field dependence.} The effective $\tilde{S}=1/2$ model uses the same bilinear exchange interaction values as in ref. \cite{neutron2021}, while the flavor wave model is a modified version of the model in ref. \cite{neutron2021}, both with external magnetic field $\bm{B}$. In the effective $\tilde{S}$=1/2 model, $\tilde{H}_{\mathrm{bl}}$ is the Hamiltonian that is bilinear on the effective $\tilde{S}=1/2$ operators with up to 6th nearest neighbor exchange interactions ($\tilde{J}_1$, $\tilde{J}_3$, and $\tilde{J}_5$ in plane and $\tilde{J}_2$, $\tilde{J}_4$, and $\tilde{J}_6$ out of plane) of the XXZ type, as shown in figure \ref{Fig1}\textbf{A}, for $\tilde{S}=1/2$ spins. $\tilde{H}_6$ is the ring-exchange interaction with strength $\tilde{\alpha}_6$, here $\tilde{\phi}_6$ represents a phase that determines the direction of the magnetic moment in the honeycomb plane and it equals \textpi. $\tilde{H}_{\mathrm{Z}}$ is the Zeeman energy of interaction between the magnetic moments and and in-plane magnetic field with an effective $g$-factor $\tilde{g}_{\parallel}$. In the flavor wave model, $H_0$ is the single Co$^{2+}$ ion Hamiltonian where $\lambda$ is the atomic spin-orbit energy, $\delta$ is the oxygen octahedron trigonal distortion energy, and $S=3/2$ and effective $l=1$. $H_{\mathrm{bl}}$ is the bilinear spin interactions and only contains isotropic (Heisenberg) nearest neighbor exchange ($J_1$ in plane and $J_4$ out of plane). $H_6$ is the ring-exchange interaction with strength $\alpha_6$ and phase $\phi_6=$\textpi. $H_{\mathrm{Z}}$ is the corresponding Zeeman energy. Uncertainties are estimated from fitting the Raman data and the neutron scattering data from \cite{neutron2021}.}
%\label{table1}
\begingroup
\def\arraystretch{1.3}
\begin{tabular}{| l @{\hspace{1ex}}  l @{\hspace{1ex}} l  @{\hspace{-0.2ex}} S[table-format=2.3]  @{\hspace{2ex}} l @{\hspace{2ex}} l  @{\hspace{1ex}} S[table-format=1.2] @{\hspace{0.5ex}} l   @{\hspace{0ex}} l || l @{\hspace{1ex}}  l @{\hspace{1ex}}   S[table-format=3.4] @{\hspace{3ex}}  l | }
\multicolumn{9}{c}{Effective $\tilde{S}=1/2$ Model} & \multicolumn{4}{c}{Flavor Wave Model} \\ \hline\hline
\multicolumn{9}{| l||}{$\tilde{H} = \tilde{H}_{\mathrm{bl}}+\tilde{H}_{6}+\tilde{H}_{\mathrm{Z}}$} & \multicolumn{4}{l|}{$H = H_0+H_{\mathrm{bl}}+H_{6}+H_{\mathrm{bq}}+H_{\mathrm{Z}}$} \\
\multicolumn{9}{|l||}{ } & \multicolumn{4}{l|}{$H_0=\sum_{\bm{r},i}(3\lambda/2) \bm{S}_{\bm{r},i}\cdot\bm{l}_{\bm{r},i}+\delta \left(\left(l^z_{\bm{r},i}\right) ^2 -2/3\right)$} \\ 
\multicolumn{9}{|l||}{$\tilde{H}_{\mathrm{bl}} = \frac{1}{2}\sum_{\bm{r},\mathrm{\text{\textdelta}} \bm{r}}\sum_{i,j}\tilde{\bm{S}}_{\bm{r},i}^\mathsf{T} \tilde{\bm{J}}_{\text{\textdelta} \bm{r}}^{ij}\tilde{\bm{S}}_{\bm{r}+\text{\textdelta} \bm{r},j}$} & \multicolumn{4}{l|}{$H_{\mathrm{bl}} = \frac{1}{2}\sum_{\bm{r},\text{\textdelta} \bm{r}}\sum_{i,j}\bm{S}_{\bm{r},i}^\mathsf{T} \bm{J}_{\text{\textdelta} \bm{r}}^{ij}\bm{S}_{\bm{r}+\text{\textdelta} \bm{r},j}$} \\ 
\multicolumn{9}{|l||}{$\tilde{H}_6=\tilde{\alpha}_{6}\bigl(\e^{-\ii\tilde{\phi}_{6}}\sum_{\bm{r}}\prod_{i=1}^{6}\tilde{S}_{\bm{r}+\bm{\delta}^{\textrm{ring}}_{i}}^+ + \textrm{h.c.}\bigr)$} & \multicolumn{4}{l|}{ $H_6=\alpha_{6}\bigl({\e}^{-\ii\phi_{6}}\sum_{\bm{r}}\prod_{i=1}^{6}S_{\bm{r}+\bm{\delta}^{\textrm{ring}}_{i}}^+ + \textrm{h.c.}\bigl)$} \\ 
\multicolumn{9}{|l||}{ } & \multicolumn{4}{l|}{$H_{\mathrm{bq}}=\frac{1}{2}\sum_{<\bm{r}i,\bm{r}'j>}q\left((S_{\bm{r}i}^+)^2(S_{\bm{r}'j}^-)^2+q^*(S_{\bm{r}i}^-)^2(S_{\bm{r}'j}^+)^2\right)$} \\  \multicolumn{9}{|l||}{$\tilde{H}_{\mathrm{Z}}=\mu_\mathrm{B}\sum_{\bm{r},i}\tilde{g}_{\parallel} \left(B^x\tilde{S}^x_{\bm{r},i}+B^y\tilde{S}^y_{\bm{r},i}\right)$} & \multicolumn{4}{l|}{$H_{\mathrm{Z}}=\mu_\mathrm{B}\sum_{\bm{r},i}\bm{B}\cdot\left(2\bm{S}_{\bm{r},i}-3\bm{l}_{\bm{r},i}/2\right)$} \\ 
\hline
\hline
$\tilde{J}_1^{xx}=$&$\tilde{J}_1^{yy}$&$=-$ & 6.36 &$\SI{}{\milli\electronvolt}$ &$\tilde{J}_1^{zz}= $ &1.97 &meV &{}&$J_{1}$ &= &-0.90(2) &meV \\
$\tilde{J}_2^{xx}=$&$\tilde{J}_2^{yy}$&$= - $ & 0.33 & meV, &$\tilde{J}_2^{zz}=$ &0.30 &meV & &{}&  & {}& \\ 
$\tilde{J}_3^{xx}=$&$\tilde{J}_3^ {yy}$&$= $ & 0.78 & meV, &$\tilde{J}_3^{zz}= $ &0.15 &meV &{}&  & & &  \\ 
$\tilde{J}_4^{xx}=$&$\tilde{J}_4^{yy}$&$=  $ & 0.11 & meV, &$\tilde{J}_4^{zz}= $ &0.32 &meV& &  $J_4$ &  $=$& 0.189(8) & meV \\ 
$\tilde{J}_5^{xx}=$&$\tilde{J}_5^{yy}$&$= - $ & 0.39 & meV, &$\tilde{J}_5^{zz}= $ &0.20 &meV& &{}& &  &   \\ 
$\tilde{J}_6^{xx}=$&$\tilde{J}_6^{yy}$&$=  $ & 0.79 & meV, &$\tilde{J}_6^{zz}= $ &0.68 &meV& &{}& && \\ 
 {} & $\tilde{g}_{\parallel}\ \ $& $=$  & 2.73(3) &   &  & &  &{}&  &  &  & \\ 
{} &  &  &  &  &  & &  &{}&$\delta$&$=$  &52(2) &meV \\ 
{} &  &   & &   &  &  & & &$\lambda$& $=$ &16.4(2) &meV \\ 
{}&$\tilde{\alpha}_6\ $ & $= $& 46(6) &$\SI{}{\micro\electronvolt}$ &  &  & & &$\alpha_6$& $=$  &0.62(7) &$\SI{}{\micro\electronvolt}$ \\
{} &  &   & &   &  &  & & &$q$& $=$ &-0.15(1) &meV \\
\hline
\end{tabular}

\endgroup
\label{table1}
\end{table*}

\section{Discussion} 

The data presented above reveal a zone-center magnon gap of $\approx\SI{0.9}{meV}$ at $\SI{8}{\kelvin}$, as measured by TDTS (figure \ref{Fig2}\textbf{B}) and $\SI{0.82(5)}{meV}$ by Raman spectroscopy (figure \ref{Fig4}\textbf{A}). A second, higher energy magnon is observed as an in-plane magnetic dipole THz absorption (figure \ref{Fig2}\textbf{A} and \textbf{C}) and is also detected in the Raman data (figure \ref{Fig4}\textbf{A}). The magnetic field dependence of these two magnons was also obtained for both in-plane and out-of-plane static magnetic fields. 

Our data confirms the existence of a magnon gap that previous theoretical models fail to adequately capture. We address this inadequacy by expanding on two models proposed in ref \cite{neutron2021}. The first model treats the magnetic moments as only coming from the lowest spin-orbital doublet of the Co$^{2+}$ ion (described above), the effective $\tilde S=1/2$ model (in ref. \cite{SI}). The second model, the flavor wave model, takes into account all six spin-orbital doublets and expresses them by bosonic operators \cite{generalizedSWT}, which are described in ref. \cite{SI}. We add to these model Hamiltonians (details in table \ref{table1}) a gap opening ring-exchange interaction that functions as an effective 6-fold anisotropy for the magnetic moments in the honeycomb plane. This generates a term in the free energy proportional to $\cos{(6\phi)}$, where $\phi$ is the angle between the spin direction and the $a$-axis. 

\begin{figure}[!t]
  \centering
  \includegraphics[width=0.45\textwidth]{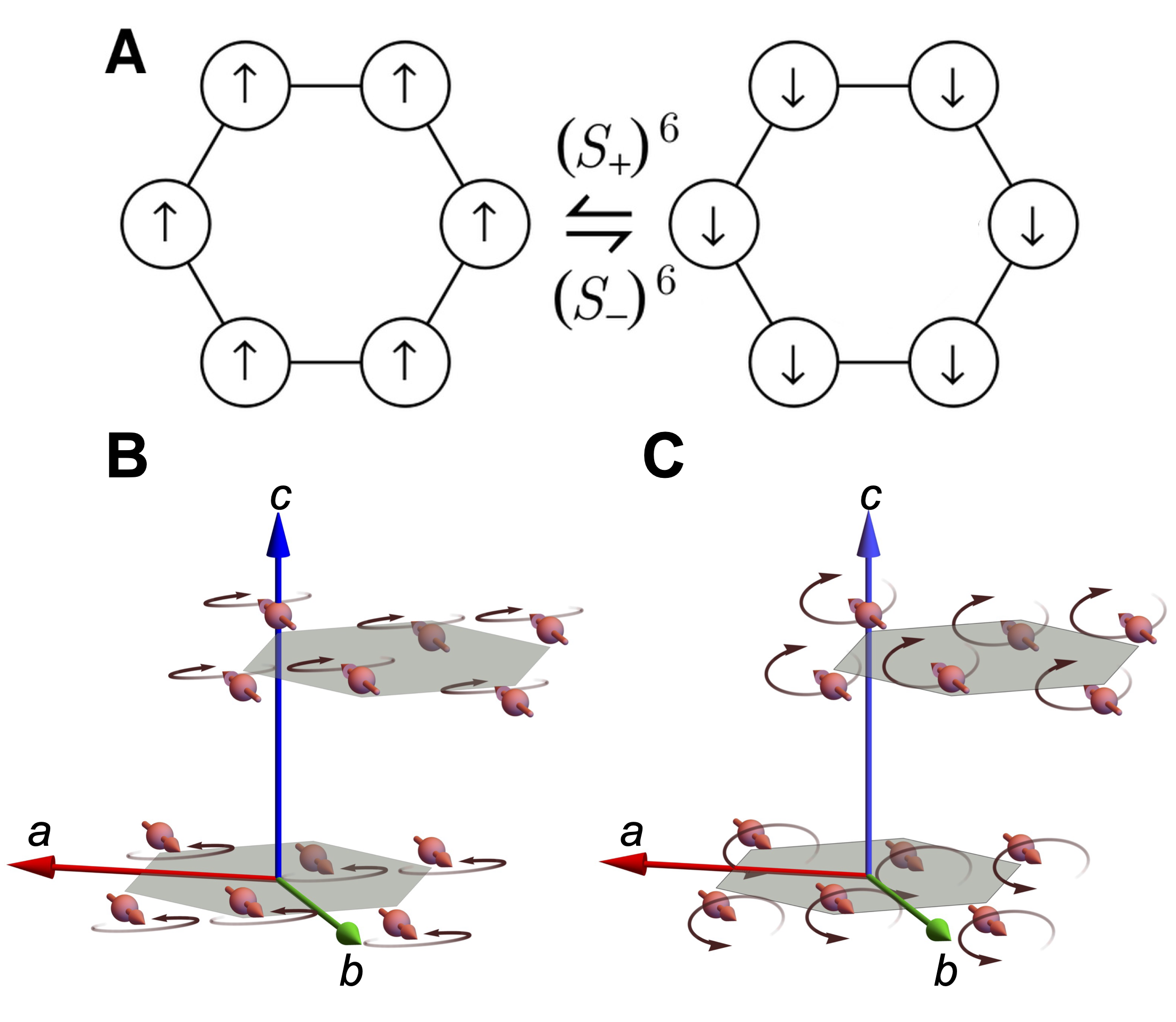}
    \caption{\textbf{Ring Exchange Interaction and Magnon Normal Modes}. \textbf{A.} Ring-exchange considered in both the $\tilde S=1/2$ and the flavor wave models. Normal modes for the pseudo-Goldstone mode and the $\SI{5.4}{\milli\electronvolt}$ mode are shown in panels \textbf{B} and \textbf{C}, respectively, where only two honeycomb layers are shown. The faded black arrows are the oscillations of the magnetic moments.}
    \label{Fig5}
\end{figure}

The form of the ring-exchange interaction we use is $r\left(\e^{-\ii\phi_{6}}\sum_{\bm{r}}\prod_{i=1}^{6}S_{\bm{r}+\bm{\delta}_{i}^{\textrm{ring}}}^{+} + \textrm{h.c.}\right)$, where $r$ is either $\tilde\alpha_6$ or $\alpha_6$ for the effective $\tilde S = 1/2$ or the flavor wave models, respectively. $\phi_6$ is an angle that determines the direction of the spins within the honeycomb plane in the ground state. This represents the simultaneous exchange of six spins within a hexagon of the honeycomb plane as schematically shown in figure \ref{Fig5}\textbf{A}. The ring-exchange interaction strength $\alpha_{6} \approx \SI{0.62}{\micro\electronvolt}$ within the flavor wave model and of $\tilde{\alpha}_{6}\approx \SI{46}{\micro\electronvolt}$ in the $\tilde{S}=1/2$ approximation can open a gap consistent with the TDTS, Raman, and neutron experiments. It is a remarkable result that such small values of the ring-exchange can open such a sizable gap of $\approx0.9$ meV, even though they are more than two orders of magnitude smaller than the gap itself and the nearest neighbor exchange interaction $J_1$. A simple estimate of the expected value of the ring exchange $\tilde{\alpha}_6$ is $6!J_1^3/U^2$, where $J_1$ is the nearest neighbor exchange interaction, $U$ is the on-site Hubbard interaction, and the factorial term is a combinatorial factor that counts how many of these ring-exchange terms contribute to the Hamiltonian. Using the value $U\approx\SI{3.25}{eV}$ from ref. \cite{DasParamekanti2021} and the nearest neighbor $J_1$ from the effective $\tilde{S}\approx 1/2$ model, we obtain $\tilde{\alpha}_6\approx \SI{17}{\micro\electronvolt}$, which is of the same order of magnitude as obtained from our fit. In the flavor wave model, we also added a biquadratic exchange between nearest neighbors with strength $q$. While this term does not open a gap, we add it to better fit the higher energy ($>\SI{24}{\milli\electronvolt}$) excitations observed in the Raman experiments \cite{thuccotio3} and in refs. \cite{neutron2021,yuan2020spin}.

The magnetic field dependence of the magnon energies (fig. \ref{Fig4}\textbf{B}) is also reproduced by these two models. In the $\tilde{S}$=1/2 model we require an effective in-plane $g$-factor $\tilde{g}_{||}=2.73(3)$ in addition to the XXZ Hamiltonian with the twelve exchange parameters on the left side of table \ref{table1}. The flavor wave model only requires the six parameters listed on the right side of table \ref{table1} to reproduce the experimentally observed gap and magnetic field dependence. We note that the magnetic field dependence is reproduced by our two models up to fields of 22 T. The details of the models are presented in ref. \cite{SI}. In the latter, we also show how it reproduces the magnetic field \cite{permeability2021}, temperature dependence of the magnetization, and the magnon dispersions \cite{yuan2020dirac,neutron2021}.

This ring-exchange term is not accessible within the expansion of the Hamiltonian bilinear in the $\tilde{S}=1/2$ operators considered in \cite{neutron2021,yuan2020dirac} which, because of its emergent $U$(1) symmetry, only has gapless excitations at the zone center. We also note that we found the flavor wave model considered in \cite{neutron2021} to be gapless even when including spin-orbital interactions. This is why we must go beyond these previous models to explain the opening of the gap. In ref. \cite{SI} we discuss other symmetry allowed terms in the flavor wave model that can open a gap. These terms, however, are not allowed within the $\tilde S= 1/2$ model, which is equivalent to the flavor wave model when only the lowest two energy levels are taken into account, and thus we do not consider them as possible explanations for the gap opening. 

In ref.\cite{neutron2021}, the finite magnon gap at the zone center was proposed to open by a quantum order-by-disorder mechanism\cite{Rau2018}. The magnon dispersion was assumed to obey a phenomenological formula $\tilde\omega({\bf k})=\sqrt{\mathit{\Delta}^2+\omega({\bf k})^2}$, $\mathit{\Delta}$ being the zone center magnon gap induced by quantum fluctuations. In contrast, by including the ring-exchange terms, we can obtain the magnon dispersion within our two models without any ad-hoc assumptions. Our theroretical results agree with data from neutron scattering (see ref. \cite{SI} for the calculated magnon dispersion), Raman spectroscopy (with magnetic field), and TDTS. Therefore, we conclude that the ring-exchange interaction is the origin of the gap at the zone center in CoTiO$_3$ .

The ring-exchange interaction also explains the selection rules associated with the absorption due to magnons in figure \ref{Fig2}. We write the normal modes of the 4 zone-center magnons, when the equilibrium spins point in the $\pm x$ direction, in the following form $\tilde S\left[(\tilde{S}^y_{1,1},\tilde{S}^z_{1,1}),(\tilde{S}^y_{1,2},\tilde{S}^z_{1,2}),(\tilde{S}^y_{2,1},\tilde{S}^z_{2,1}),(\tilde{S}^y_{2,2},\tilde{S}^z_{2,2})\right]$, where $\tilde{S}^\beta_{\mu,\nu}$ is the oscillating part of the spin along the $\beta$-axis, in layer $\mu$ and sublattice $\nu$. Without the ring exchange, the lowest energy mode would be gapless, i.e. a Goldstone mode due to $U(1)$ symmetry in the XXZ model. Its normal mode, within the $\tilde S=1/2$ approximation, is $\tilde S\left[(0.5,0),(0.5,0),(-0.5,0),(-0.5,0)\right]$. This means that the spins only deviate from the equilibrium direction in the hexagonal plane and have no net magnetic moment associated with it because $\sum_{\mu,\nu}\tilde{S}^\beta_{\mu,\nu}=0$ for both $\beta=y$ and $z$. However, with the ring-exchange interaction, the normal mode becomes $\tilde S\left[(0.5,0.03\ii),(0.5,0.03\ii),(-0.5,0.03\ii),(-0.5,0.03\ii)\right]$ as shown in fig. \ref{Fig5}\textbf{B}. Thus, this magnon has a net magnetic dipole moment along the $c$-axis ($\sum_{\mu,\nu}\tilde{S}^z_{\mu,\nu}\neq 0$), consistent with the TDTS data shown in figure \ref{Fig2}\textbf{B}, where the absorption due to this pseudo-Goldstone mode happens only when $\bm{h}_\omega\parallel c$-axis. 

The normal mode for the 5.4 meV magnon does not qualitatively change with the addition of the ring-exchange and keeps its net magnetic moment in the honeycomb plane. Thus, it can be excited by $\bm{h}_\omega\parallel$ to the honeycomb plane, i.e. $\sum_{\mu,\nu}\tilde{S}^y_{\mu,\nu}\neq 0$ as shown in figures \ref{Fig2}\textbf{A} and \textbf{C}. Its normal mode, shown in fig. \ref{Fig5}\textbf{C}, is $\tilde S\left[(0.45,0.23\ii),(0.45,0.23\ii),(0.45,-0.23\ii),(0.45,-0.23\ii)\right]$. This mode is the one whose energy decreases toward zero around $\SI{17}{\tesla}$ as shown in figure \ref{Fig4}\textbf{B}. Its intensity vanishes in the fully polarized state as a result of the magnon being at the Brillouin zone edge along the $k_z$ direction, which is thus Raman inactive. We can tell this is the case because its normal mode has $y$-axis spin oscillations of opposite sign for the 2 layers along the $c$-axis making its periodicity (two layers) twice the unit cell (one layer) in the fully polarized state. The details of the normal modes of the other two zone-center magnons are given in ref. \cite{SI}.

Finally, we address the topic of topologically protected surface magnons in CoTiO$_3$ by analyzing the symmetry indicators \cite{Po2017,topologicalquantumchemistry,lu2022} of the magnon wavefunctions within the $\tilde{S}=1/2$ model. The symmetry indicators of the bond-centered inversion symmetry ($\mathcal{I}$) of the two lower magnon modes imply that, when there is a gap between them in the entire Brillouin zone, both magnon bands must have odd Chern numbers. The magnetic inversion symmetry $\tilde{\mathcal{I}}$ (defined in ref. \cite{SI}), however, forces the Chern number to vanish when magnons are energetically separated from each other. Because the Chern number cannot be both an odd integer and zero simultaneously, these magnon bands must cross somewhere in the Brillouin zone. Experiments show that they intersect at two nodal lines around the hexagonal Brillouin zone corners (K and K' points) \cite{yuan2020dirac,neutron2021} and are reproduced here with our modeling  \cite{SI}. These nodal lines are topologically protected by the magnetic inversion $\tilde{\mathcal{I}}$ and are characterized by a non-trivial first Stiefel-Whitney number \cite{Ahn2019}. However, this symmetry is generally broken by an open surface, and thus CoTiO$_3$ cannot have protected topological magnons on its surfaces.

\section{Conclusions}

In summary, with the combination of TDTS, magneto-Raman spectroscopy, and theoretical modeling we determined that the origin of the magnon gap in CoTiO$_3$ is a ring-exchange interaction between the magnetic moments in the honeycomb plane. We found that this interaction acts the same way in both the flavor wave model and the effective $\tilde S=1/2$ approximation. By applying a magnetic field along the honeycomb plane, we determined the field dependence of the two lowest magnon energies. This allowed us to refine the values of the exchange interactions in the Hamiltonians used to model the experiment. A surprisingly small value for the ring-exchange is responsible for the $\approx\SI{0.9}{meV}$ magnon gap. The ring-exchange interaction also explains the selection rules for magnon absorption in the TDTS experiment. It is an exciting possibility that the ring-exchange plays a significant role in the physics of other quantum magnets. Furthermore, we showed that no protected topological magnons can occur on the surface of CoTiO$_3$. More generally, this work demonstrates the power of combining TDTS, magneto-Raman spectroscopy, and theoretical modelling to study magnon excitations in quantum magnets.

\section{Acknowledgements}

Work at OSU and CSU was supported by the Center for Emergent Materials at OSU, a Materials Research Science and Engineering Center funded by NSF under grant DMR-2011876. AJB and ARHW acknowledge the NIST Innovation in Measurement Science (IMS) program for funding. The authors declare that they do not have any competing financial interests. The identification of any commercial product or trade name does not imply endorsement or recommendation by the National Institute of Standards and Technology. 

Contributions: YL, EVJ, DMH, CL, and AG built the TDTS system, obtained, and analyzed the TDTS data at OSU. TTM, JRS, TA, and MFM obtained and analyzed the Raman data at NIST. AP and CF obtained the Raman data at LNCMI. YL, MK, JG, YML, NT, and KFG performed theoretical analysis and calculations to explain the magnon spectrum and its magnetic field dependence. TDL, DS, and KR grew, cut, and oriented the single crystals of CoTiO$_3$ used in this work. AJB and RLD performed magnetization measurements and additional Laue x-ray diffraction on the single crystals. RVA and YL wrote the manuscript with input from all the coauthors. RVA generated the initial idea for this project. ARHW, NT, YML, and RVA supervised all the work.

\appendix

\bibliography{CTO_Biblio}% Produces the bibliography via BibTeX.

\end{document}

\title{Supplemental Material: Ring-Exchange Interaction Effects on Magnons in Dirac Magnet CoTiO$_3$}% 
\author{Yufei Li}
\affiliation{Department of Physics, The Ohio State University. Columbus, OH 43210}
\author{Thuc T. Mai}%
\affiliation{Quantum Measurement Division, Physical Measurement Laboratory, NIST, Gaithersburg, MD 20899}
\author{M. Karaki}%
\affiliation{Department of Physics, The Ohio State University. Columbus, OH 43210}
\author{E.V. Jasper}%
\affiliation{Department of Physics, The Ohio State University. Columbus, OH 43210}
\author{K.F. Garrity}%
\affiliation{Materials Measurement Science Division, Materials Measurement Laboratory, NIST, Gaithersburg, MD 20899}
\author{C. Lyon}%
\affiliation{Department of Physics, The Ohio State University. Columbus, OH 43210}
\author{D. Shaw}%
\affiliation{Department of Physics, Colorado State University, Fort Collins, CO 80523}
\author{T. DeLazzer}%
\affiliation{Department of Physics, Colorado State University, Fort Collins, CO 80523}
\author{A.J. Biacchi}
\affiliation{Nanoscale Device Characterization Division, Physical Measurement Laboratory, NIST, Gaithersburg, MD 20899}
\author{R.L. Dally}
\affiliation{NIST Center for Neutron Research, National Institute of Standards and Technology, Gaithersburg, MD 20899}
\author{D.M. Heligman}%
\affiliation{Department of Physics, The Ohio State University. Columbus, OH 43210}
\author{J. Gdanski}%
\affiliation{Department of Physics, The Ohio State University. Columbus, OH 43210}
\author{T. Adel}%
\affiliation{Quantum Measurement Division, Physical Measurement Laboratory, NIST, Gaithersburg, MD 20899}
\author{M.F. Mu\~noz}%
\affiliation{Quantum Measurement Division, Physical Measurement Laboratory, NIST, Gaithersburg, MD 20899}
\author{A. Giovannone}%
\affiliation{Department of Physics, The Ohio State University. Columbus, OH 43210}
\author{A. Pawbake}
\affiliation{Laboratoire National des Champs Magnetiques Intenses, LNCMI-EMFL. Grenoble, France}
\author{C. Faugeras}
\affiliation{Laboratoire National des Champs Magnetiques Intenses, LNCMI-EMFL. Grenoble, France}
\author{J.R. Simpson}%
\affiliation{Department of Physics, Astronomy, and Geosciences, Towson University, Towson, MD 21252}
\affiliation{Quantum Measurement Division, Physical Measurement Laboratory, NIST, Gaithersburg, MD 20899}
\author{K. Ross}%
\affiliation{Department of Physics, Colorado State University, Fort Collins, CO 80523}
\author{N. Trivedi}%
\affiliation{Department of Physics, The Ohio State University. Columbus, OH 43210}
\author{Y.M. Lu}%
\affiliation{Department of Physics, The Ohio State University. Columbus, OH 43210}
\author{A.R. Hight Walker}%
\affiliation{Quantum Measurement Division, Physical Measurement Laboratory, NIST, Gaithersburg, MD 20899}
\author{R. Vald\'es Aguilar}
\email{valdesaguilar.1@osu.edu}
\affiliation{Department of Physics, The Ohio State University. Columbus, OH 43210}

\date{\today}% It is always \today, today,

\renewcommand{\thetable}{S\arabic{table}}  
\renewcommand{\thefigure}{S\arabic{figure}}
\renewcommand{\theequation}{S.\arabic{equation}}

\renewcommand{\bibnumfmt}[1]{[S#1]} % This line requires natbib
\renewcommand{\citenumfont}[1]{S#1} % This line requires natbib

\maketitle
\section{Supplementary Note 1: Single crystal growth and characterization}
\begin{figure}[hb]
  \centering
    \includegraphics[width=.35\textwidth]{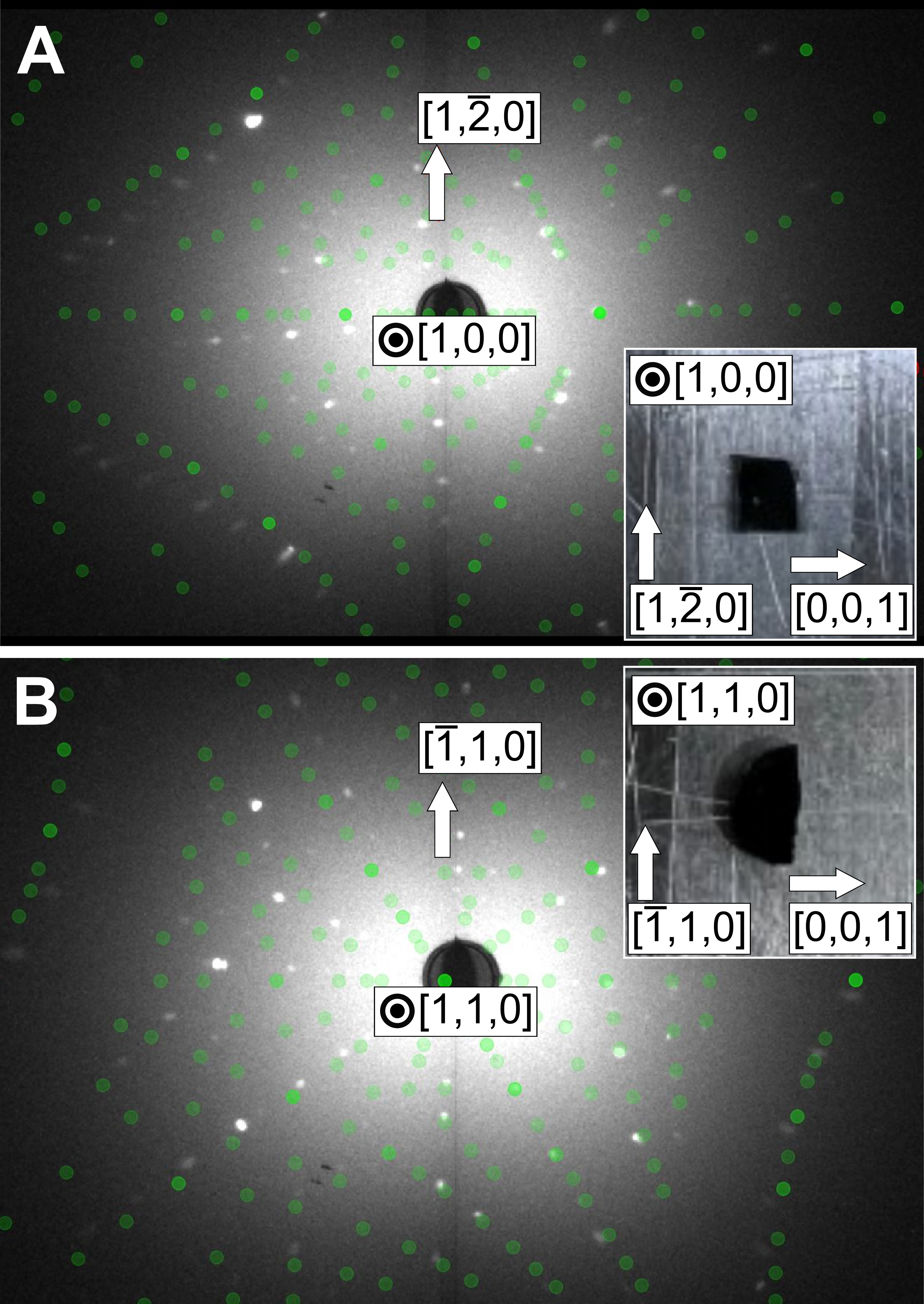}
    \caption{Laue diffraction patterns for crystals cut perpendicular to the \textbf{A} [100] and \textbf{B} [110] axis. Insets show pictures of both crystals.}
    \label{FigLaue}
\end{figure}

Single crystals were grown using solid state synthesis, starting rods were made by reaction of a stoichiometric mixture of Co$_3$O$_4$ and TiO$_2$. The rods were sintered three times at 1050 {$^\circ$C} for 14 to 18 hours with intermediate regrinding. A crystal was grown by the optical floating zone method using a growth rate of $\SI{5}{\milli\meter/\hour}$, in an 1 atmosphere static air environment. From this crystal several pieces were obtained and oriented perpendicular to the [001] and [110] axes. A Photonic Science X-Ray Laue Diffractometer ($\SI{40}{\volt}$, $\SI{400}{\micro\ampere}$) was used to obtain orientation information on two CoTiO$_3$ single crystals, shown in Fig. \ref{FigLaue} \textbf{A} and \textbf{B}.

Field-cooled DC magnetization measurements were carried out using a Quantum Design Physical Property Measurement System (PPMS) equipped with a $\SI{12}{\tesla}$ superconducting magnet on the two crystals shown in figure \ref{FigLaue}.  CoTiO$_3$ single crystals were mounted onto a quartz paddle using VGE-7031 varnish (from Lakeshore) with different crystallographic axis parallel to the applied magnetic field.  Crystals were removed from the paddle for reorientation by sonicating in acetone for three hours.  The PPMS vibrating sample magnetometer (VSM) module was employed and the chamber was evacuated to a pressure of $\SI{5}{Torr}$($\SI{0.66}{\kilo\pascal}$) or less.  Upon heating to $\SI{400}{\kelvin}$, a $\SI[retain-explicit-plus]{+0.1}{\tesla}$ field was applied.  The temperature was then lowered at a rate of $\SI{2}{\kelvin/\minute}$ and mass-normalized magnetization was measured down to $\SI{2}{\kelvin}$.

\begin{figure}[b]
  \centering
    \includegraphics[width=.65\textwidth]{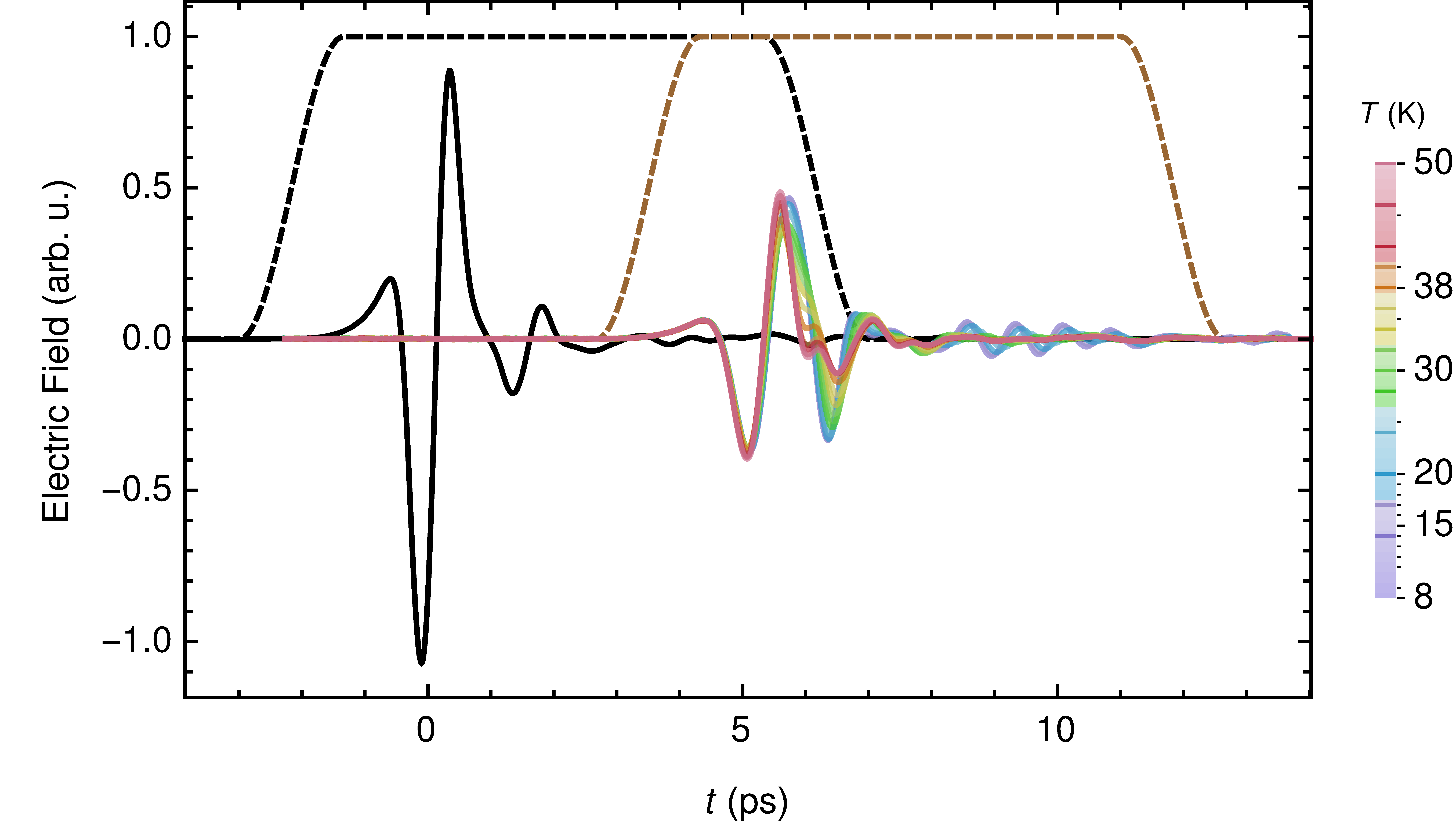}
    \caption{Time-domain THz data of aperture (black line) at 8 K and sample below 50 K. Sample temperatures are indicated in color bar. Black and brown dashed lines are the window functions applied to aperture and sample traces, respectively, before Fourier transformation. Sample is $\sim\SI{1}{\milli\meter}$ thick.}
    \label{FigSuppl1}
\end{figure}

\section{Supplementary Note 2: THz data}
Terahertz measurements were performed at OSU using a home-built time Domain Terahertz spectroscopy (TDTS) system. Two methods of transmission TDTS were used: an antenna based system and an another based on terahertz spintronic emitters\cite{spintronicemitter}, both yielding the same transmission results. The antenna based system has been described elsewhere \cite{ThucSFSO,MWarrenGVS,EvanFresnelRhomb}. The spintronic TDTS system is based on a $\SI{800}{\nano\meter}$ titanium:sapphire amplified ultrafast laser with $\SI{1}{\kilo\hertz}$ repetition rate emitting pulses of approximately $\SI{60}{\femto\second}$  in duration. The generated THz pulse from the spintronic emitter was collimated and one $\ang{90}$ off-axis parabolic mirror focuses the THz pulses onto the sample which sits inside a closed-cycle cryostat that can reach temperatures as low as $\SI{8}{\kelvin}$. The detection of the transmitted THz pulse is done via standard electro-optic sampling using a thin GaSe nonlinear optical crystal\cite{GaSedetection}. The transmitted THz pulse is Fourier transformed to obtain its spectrum and referenced to the spectrum obtained from an empty aperture in place of the sample. The raw pulses of Fig. 2A from main text, and the window functions used are shown in figure \ref{FigSuppl1}. Time-domain data are multiplied by window functions and then Fourier transformed to get the spectra in frequency domain. The ratio of the magnitudes of these spectra generates the transmission function of the sample. The linear polarization of the THz pulses is controlled using THz wire grid polarizers. Several crystals of CoTiO$_3$ cut perpendicular to specific crystalline axes were used for THz experiments in order to determine the linear response of the magnetic excitations. 

\section{Supplementary Note 3: Raman data}
\begin{figure}[b]
  \centering
    \includegraphics[width=0.55\textwidth]{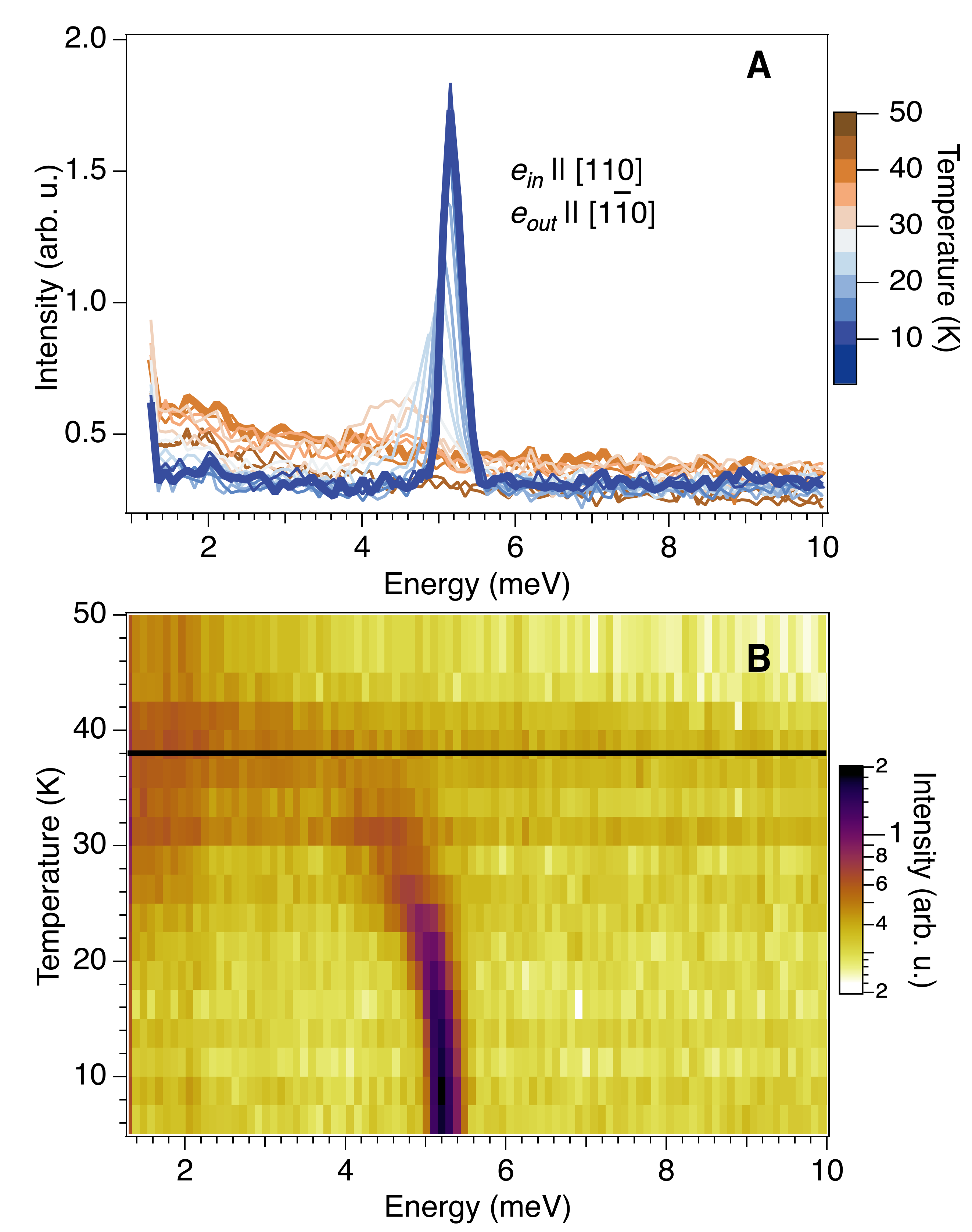}
    \caption{\textbf{Temperature dependence of Raman spectrum.} \textbf{A}. Raman spectra of the low energy region for several temperatures below $\SI{50}{\kelvin}$ in the VH configuration. \textbf{B}. False color map of the Raman data shown in \textbf{A}. Black horizontal line indicates the N\'eel temperature of $\SI{38}{\kelvin}$.}
    \label{FigSuppl2}
\end{figure}

\begin{table*}[t] 
\begingroup
\begin{tabular}{| c | c |c|c|c|}
\hline
Magnetic Field(T) & Mode 1(meV) & FWHM 1(meV) & Mode 1(meV) & FWHM 2(meV)\\
\hline
0     & 0.82(5)   &   0.22  &  5.37(5)    &  0.23  \\ \hline
3     & 1.28(5)   &   0.32  &   5.31(5)   &   0.24 \\ \hline
6     & 2.17(5)   &     0.22    &   4.98(5)   &   0.32 \\ \hline
9     & 3.18(5)   &    0.18     &  4.44(5)    &   0.25 \\ \hline
\end{tabular}
\endgroup
\caption{Center frequencies and full-widths at half maxima (FWHM) extracted from a Gaussian fit to the Raman data for selected magnetic fields. Magnetic field is perpendicular to $c$-axis.}
\label{RamanDataTable}
\end{table*}

\begin{figure}[hb]
  \centering    \includegraphics[width=0.85\textwidth]{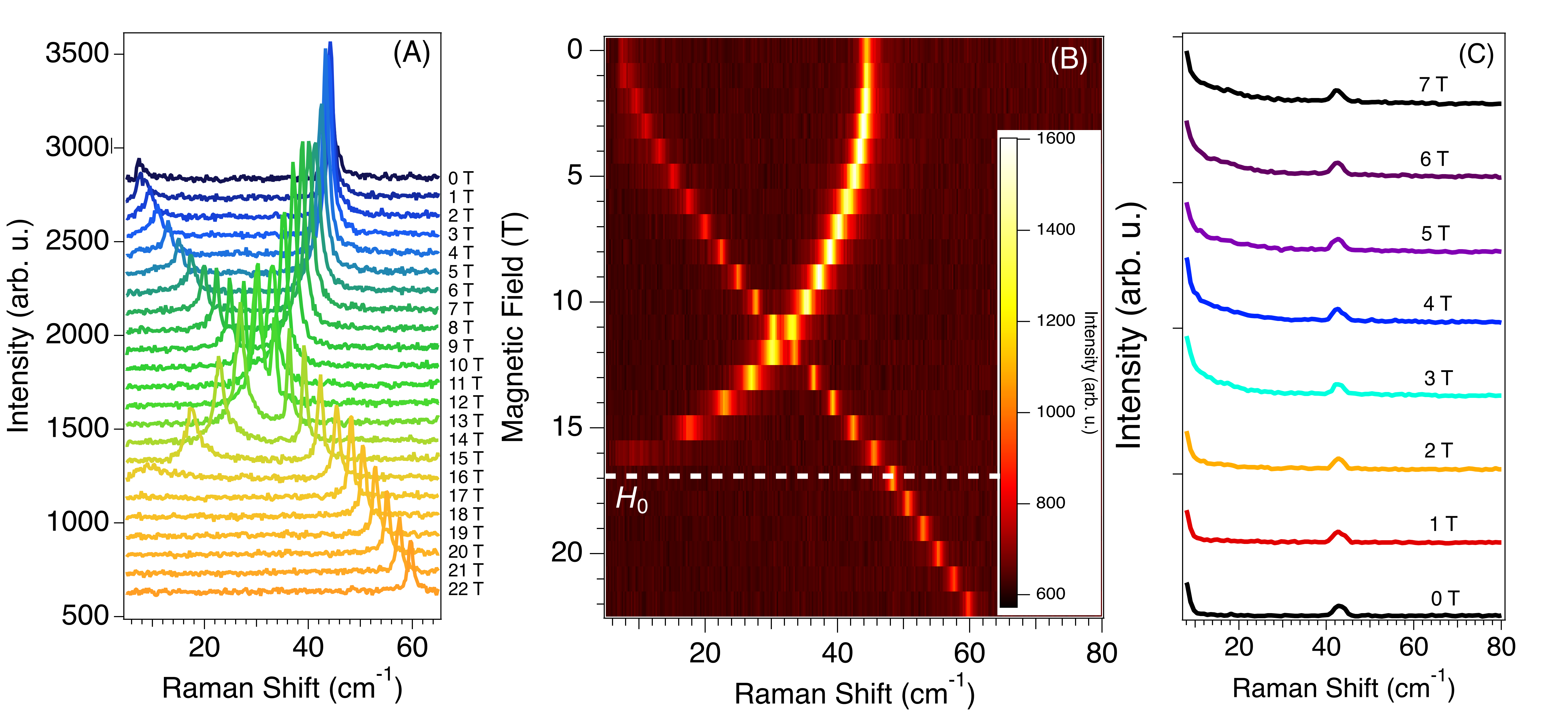}
    \caption{\textbf{High Magnetic Field dependence of Raman spectrum.} (A) Unpolarized Raman sepctra at a temperature of 5 K up to 22 T applied along the [110] crystallographic direction. (B) False color intensity map of the data shown in (A). (C) Magnetic field dependence of 5.4 meV magnon for $H$ parallel to the [001] direction.}
    \label{FigSuppl3.5}
\end{figure}

The Raman spectroscopy measurements in a magnetic field were performed at NIST using a triple-grating spectrometer that has been used earlier \cite{NISTRaman2, McCrearyFePS3} to study magnetic materials. In this work, the $c$-cut single crystal (the surface of the sample is perpendicular to the [001] direction) was mounted inside a cryogen-free magneto-cryostat with optical access. The sample space is filled with Helium as thermal exchange gas. The sample temperature can be varied from $\SI{1.6}{\kelvin}$ to $\SI{300}{\kelvin}$, and an external magnetic field up to $\SI{9}{\tesla}$ can be applied to the sample. The sample is placed in the Voigt geometry, where the sample surface is parallel to the magnetic field's direction. Two different gas lasers are used, $\SI{633}{\nano\meter}$ (HeNe) and $\SI{488}{\nano\meter}$ (Ar$^+$), and detect the same two magnon excitations. All optical beams are directed to and collected from the sample via free-space optics. The incoming laser beam is at normal incidence to the sample through a focusing optic. The Raman back-scattered beam is collected and sent to a triple grating spectrometer. Various polarization optics are in the beam path to control the incident and scattered beam polarizations. 

Two main polarization configurations are used: parallel (VV) with incoming and scattered polarization parallel to each other; and crossed (VH) with incoming and scattered polarization perpendicular to each other. The laser spot size on the sample is approximately $\SI{1}{\micro\meter}$. The uncertainty of the center frequency, combined from the uncertainty of a Gaussian fit and systematic uncertainty, is mostly from the pixel width ($\SI{0.4}{\centi\meter^{-1}}=\SI{0.05}{\milli\electronvolt}$) of the spectrometer. The instrument broadening (FWHM) is estimated to be $\SI{1.1}{\centi\meter^{-1}}=\SI{0.14}{\milli\electronvolt}$. The temperature dependence of the VH Raman spectrum is shown in figure \ref{FigSuppl2}. Here we show spectra below $\SI{50}{\kelvin}$ and in the energy range below $\SI{10}{\milli\electronvolt}$ where only one mode is observed. The pseudo-Goldstone mode was measured only when the magnetic field dependence was studied as shown in the main text. Table \ref{RamanDataTable} shows values of the magnon energies and FWHM at selected values of the applied magnetic field.

Additional Raman scattering measurements up to 22 T were performed at the high magnetic field laboratory (LNCMI) in Grenoble, France. Unpolarized light of wavelength 633 nm was used in a backscattering configuration, where the incident light propagates at an angle of $\sim 30^\circ$ from the normal of the sample surface. The sample was at a base temperature of 5 K in a He gas environment. The magnetic field was applied along the [110] crystallographic direction. These data are shown in figure \ref{FigSuppl3.5}.

\section{Supplementary Note 4: Effective $\tilde{S}$= 1/2 model}
In this section, we discuss the $\tilde{S}$= 1/2 model, which has been used in \cite{yuan2020dirac,neutron2021}. Co$^{2+}$ ions in CoTiO$_3$ form a layered honeycomb lattice with a A-B-C stacking between the two layers. More precisely, the Bravais lattice vectors are given by
\bea
\bm{a}_1=\bm{a},~~~\bm{a}_2=\bm{b},~~~\bm{a}_3=\frac{-\bm{a}-2\bm{b}+\bm{c}}{3}.
\eea
where
\bea
\bm{a}=\sqrt3 (1,0,0)a,~~~\bm{b}=(-\sqrt3,3,0)a/2,~~~\bm{c}=(0,0,c).
\eea
are lattice basis vectors. Here $a=\SI{5.06383(3)}{\angstrom}$ and $c=\SI{13.9076(1)}{\angstrom}$\cite{neutron2021}.
The two sublattices of the honeycomb lattice, labeled by sublattice index $s=\pm$, are coordinated at
\bea
\bm{r}_{\pm}=\pm(0,a/2,0).
\eea
Below we label a lattice site by its coordinates $(x_1,x_2,x_3,s)$ as
\bea
\bm{R}(x_1,x_2,x_3,s)=\sum_{i=1}^3x_i\bm{a}_i+\bm {r}_s
\eea

The space group is $R\bar3$, generated by site-centered (or hexagon centered) 3-fold rotation $C_3$, and bond-centered inversion $\mathcal{I}$. Below we consider a few dominant exchange couplings suggested by neutron scattering studies \cite{yuan2020dirac,neutron2021}.

The in-plane 1st nearest-neighbor (NN) exchange coupling
\bea
\hat J_1\equiv\langle0,0,0,+|0,0,0,-\rangle
\eea
is constrained by inversion symmetry only, and has the following generic form
\bea
\hat J_1=\hat J_1^{\mathsf{T}}=\bpm J_1&\Gamma_{1,xy}&\Gamma_{1,xz}\\\Gamma_{1,xy}&J_1+K_1&\Gamma_{1,yz}\\ \Gamma_{1,xz}&\Gamma_{1,yz}&J_1+J_{1,z}\epm
\eea
where $J_1,K_1,J_{1,z}$ stand for Heisenberg, Kitaev and Ising type interactions respectively. Meanwhile, $\Gamma$'s are the symmetric anisotropic exchange couplings.

The in-plane 2nd NN exchange coupling is
\bea
\hat J_3\equiv\langle0,0,0,+|1,0,0,+\rangle
\eea
It has no constraint and is an arbitrary real $3\times3$ matrix by symmetry.

The dominating inter-plane exchange couplings have two inequivalent terms for two sublattices $s=\pm$
\bea
\hat J_{4,s}\equiv\langle0,0,0,s|0,0,1,s\rangle
\eea
Subject to no symmetry constraints, they can also be any $3\times 3$ real matrices.

\subsection*{Magnon wavefunctions and symmetry representations}

The magnetic order is A-type antiferromagnetic (AFM), meaning ferromagnetic within each honeycomb plane while two neighboring planes are antiferromagnetically aligned. The ordered moments have the following form:
\bea
\expval{\bm{S}(x_i,s)}\propto(-1)^{x_3}(1,0,0)
\eea
This magnetic order breaks the $C_3$ symmetry as well as the Bravais translation $T_3$ along $\bm{a}_3$, but preserves the link-centered inversion $\mathcal{I}$, as well as magnetic translation $\tilde T_3\equiv T_3\cdot\bst$ where $\bst$ is the time reversal symmetry. The combination of inversion and magnetic translation is a magnetic inversion symmetry
\bea
\tilde{\mathcal{I}}=\mathcal{I}\cdot T_3\cdot\bst
\eea
which forces the Berry curvatures of any magnon wavefunctions to vanish identically except for at the touching points of two (or more) magnon bands. Expanding the spin variables around the ordering moments, the low-energy spin wave variables are
\bea\label{lsw basis}
\bm{\phi}_{(x_1,x_2,x_3)}\equiv(S^y_{(x_1,x_2,2x_3,s)},S^z_{(x_1,x_2,2x_3,s)},S^y_{(x_1,x_2,2x_3+1,s)},S^z_{(x_1,x_2,2x_3+1,s)})^{\mathsf{T}}.
\eea
in the doubled magnetic cell. And in momentum space it transforms under symmetry as
\bea
&\bm{\phi}_{\bm k}\overset{\mathcal{I}}\longrightarrow\bpm1&0\\0&e^{-\imth k_3}\epm_{\mu}\otimes\tau_x\otimes\sigma_0\bm{\phi}_{-\bm k},\\
&\bm{\phi}_{\bm k}\overset{\tilde T_3}\longrightarrow\bpm0&1\\e^{-\imth k_3}&0\epm_{\mu}\otimes\tau_0\otimes(-\sigma_0)\bm{\phi}_{-\bm k}.
\eea
where $\bm\mu$,~$\bm\tau$ and $\bm\sigma$ are Pauli matrices for the layer, sublattice and spin component indices, respectively. In a realistic model dominated by ferromagnetic $J_1$ and antiferromagnetic $J_4$, a direct calculation shows the two lower magnon branches at zone center $\bm k=(0,0,0)$ are both even under inversion, and their wavefunctions have the following form
\bea\label{magnon wf:optical}
b_{\bm k=\bm0}(\mathcal{I}=+1)\sim\sum_{s,\mu=\pm}S^y_{\bm k=\bm 0,s,\mu}+(\alpha+\imth\mu\beta)S^z_{\bm k=\bm 0,s,\mu},~~~\alpha,\beta\in\mathbb{R},~~~\tilde T_3=-1.
\eea
or
\bea\label{magnon wf:acoustic}
b_{\bm k=\bm0}(\mathcal{I}=+1)\sim\sum_{s,\mu=\pm}\mu S^y_{\bm k=\bm 0,s,\mu}+(\mu\alpha+\imth\beta)S^z_{\bm k=\bm 0,s,\mu},~~~\alpha,\beta\in\mathbb{R},~~~\tilde T_3=+1.
\eea
where $s,\mu=\pm$ label the sublattice and layer indices respectively.

In previous INS studies \cite{yuan2020dirac,neutron2021}, an extended XXZ model is obtained by fitting the magnon spectrum observed in INS. The XXZ model has a higher symmetry than the most generic model presented above, in the sense that there is an additional $U(1)_z\rtimes Z_2^x$ global spin rotational symmetry in addition to $R\bar3$ space group symmetries. This enlarged symmetry group is broken down to a $Z_2^x\times Z_2^{z\cdot\bst}$ symmetry by the magnetic order, generated by
\bea
&\bm{\phi}_{\bm k}\overset{e^{\imth\text{\textpi} S^x}}\longrightarrow\mu_0\otimes\tau_0\otimes(-\sigma_0)\bm{\phi}_{\bm k},\\
&\bm{\phi}_{\bm k}\overset{e^{\imth\text{\textpi} S^z}\cdot\bst}\longrightarrow\mu_0\otimes\tau_0\otimes\sigma_z\bm{\phi}_{-\bm k}
\eea
These extra symmetries dictate that $\alpha=0$ in (\ref{magnon wf:optical})-(\ref{magnon wf:acoustic}). This determines the form of wavefunctions for all inversion-even magnon modes at zone center.

Specifically, in the XXZ model, the lowest energy soft mode involves only the easy-plane variable $S^{x,y}$, and hence the lowest energy (``pesudo-Goldstone'') magnon mode in the XXZ model is nothing but a special case of (\ref{magnon wf:acoustic}):
\bea\label{magnon wf:goldstone}
b_{\bm k=\bm0}(\text{pseudo-Goldstone},\mathcal{I}=+1)\sim\sum_{s,\mu=\pm}\mu\cdot S^y_{\bm k=\bm 0,s,\mu}
\eea
The 2nd lowest magnon above the pseudo-Goldstone mode has a wavefunction of the form (\ref{magnon wf:optical}).

\subsection*{Gap opening of the Goldstone mode}

In the XXZ model, due to the spontaneous breaking of $U(1)_{\tilde{S}^z}$ spin rotational symmetry in the in-plane AFM ground state, a linearly dispersing gapless magnon mode (\ie the Goldstone mode) will appear at the zone center. Na\"ively, this Goldstone mode can be gapped out by anisotropic exchange couplings. To be specific, the linear spin wave Hamiltonian in the basis (\ref{lsw basis}) writes
\bea
\mathcal{H}_\text{LSW}=\sum_{\bm{k}}\bm{\phi}_{-\bm{k}}^{\mathsf{T}}\bm{R}(k)\bm{\phi}_{\bm{k}},~~~\bm{R}^\ast({\bm{k}})=\bm{R}^{\mathsf{T}}({\bm{k}})=\bm{R}(-{\bm{k}}).
\eea
where $\bm{R}(\bm{k})$ is a $8\times8$ positive-definite matrix. The Schr\"odinger equation for spin wave variable $\bm{\phi}_{\bm{k}}$ writes
\bea
\imth\frac{\text{d}}{\text{d}t}\bm{\phi}_{\bm{k}}=\tilde{S}\bm{Y}\cdot \bm{R}(\bm{k})\bm{\phi}_{\bm{k}},~~~\bm Y\equiv\mu_z\otimes\tau_0\otimes\sigma_y.
\eea
The linear spin wave (LSW) spectrum generally have a particle-hole symmetry in the sense that eigenstates $\bm{v}_k$ with energy $\omega(\bm{k})$ and $\bm{v}_{-\bm{k}}=\bm{v}_{\bm{k}}^\ast$ with energy $\omega(-\bm{k})=-\omega(\bm{k})$ always appear in pairs. Focusing on the zone center $\bm k=(0,0,0)$, the Goldstone mode (\ref{magnon wf:goldstone}) appears at zero energy in the XXZ model. What are the effects of anisotropic exchange interactions in the LSW theory?

For a (pseudo)spin-$1/2$ system based on Co$^{2+}$ ions, there are only bilinear exchange couplings $J_{ij}^{ab}\tilde{S}_i^a\tilde{S}_j^b$ between two spins at sites $i,j$. Note that the in-plane spin components transform nontrivially under the $C_3$ space group symmetry
\bea
\tilde{S}_{\bm r}^\pm\overset{C_3}\longrightarrow \e^{\pm\imth2\text{\textpi}/3}\tilde{S}_{C_3{\bm r}}^\pm
\eea
This means the discrete $C_3$ symmetry is in fact promoted to a continuous $U(1)$ symmetry at zone center $\bm k=(0,0,0)$. Therefore the Goldstone mode (\ref{magnon wf:goldstone}) is robustly gapless within the LSW theory, as long as $C_3$ symmetry is preserved.

The only way to gap out the Goldstone mode (\ref{magnon wf:goldstone}) is to break the enlarged $U(1)$ symmetry at $\bm k=0$ in the LSW theory of a bilinear spin model. To do so, we need to introduce $C_3$-symmetric multi-spin interactions beyond bilinear terms in the Hamiltonian, such as quartic terms involving four different spin-$\frac12$'s. For example, the following quartic term
\bea
-\imth(\prod_{\bm\delta\in\text{N.N.}}\tilde{S}^+_{\bm r+\bm\delta})\tilde{S}^z_{\bm r}=\tilde{S}^2\sum_{\bm\delta}s^y_{\bm r+\bm\delta}s^z_{\bm r},~~~\bm{S}_{\bm{ r}}=\big(\tilde{S}-\frac{(s^y_{\bm r})^2+(s^z_{\bm r})^2}{2\tilde{S}},s_{\bm r}^y,s_{\bm r}^z\big)+O(|\bm s_{\bm r}|^4)
\eea
preserves the $C_3$ symmetry, where $\bm\delta$ are vectors connecting nearest neighbor pairs. However, this quartic interaction would change the ground state by giving the spins a [001] component staggered between the honeycomb planes. Therefore, we do not consider this interaction any more. A symmetric six-spin ring-exchange interaction can give rise to the following terms:
\bea
\prod_{i=1}^6\tilde{S}^+_{\bm {r}_i}=-\tilde{S}^4\big(\sum_i\frac{(s^y_{\bm r_i})^2+(s^z_{\bm r_i})^2}2+\sum_{i,j}s_{\bm r_i}^ys_{\bm r_j}^y\big)+O(|\bm s_{\bm r}|^4)
\eea
These multi-spin terms clearly break the $C_3$ rotational symmetry in the LSW theory, and can induce a finite gap for the Goldstone mode (\ref{magnon wf:goldstone}).

\subsection*{Spin model and classical ground state}
Consider the Hamiltonian with a magnetic field in the $ab$ plane
\begin{align*}
  H &= H_{\textrm{bl}}+H_{6}+H_{\textrm{Z}}\\
   &= 
   \sum_{\bm{r}, \updelta \bm{r}}\sum_{i,j}\tilde{\bm{S}}_{\bm{r},i}^\mathsf{T} \bm{J}_{\updelta \bm{r}}^{ij}\tilde{\bm{S}}_{\bm{r}+\updelta \bm{r},j}
     +\tilde{\alpha}_{6}\bigl(\e^{-\ii\tilde{\phi}_{6}}\sum_{\bm{r}}\prod_{i=1}^{6}\tilde{S}_{\bm{r}+\bm{\delta}^{\textrm{ring}}_{i}}^+ + \textrm{h.c.}\bigr)\\
     &+ \frac{g_{\parallel}\mu_{\textrm{B}}B_{\parallel}}{2}\sum_{\bm{r}}\sum_{i}\bigl(\e^{-\ii\tilde{\phi}_{b}}\bm{S}_{\bm{r},i}^+ +\textrm{h.c.}),
\end{align*}
where $\tilde{\phi}_{b}$ is the angle between the in-plane magnetic field and crystal axis $\bm{a}=a\hat{\bm{x}}$.

In magnetic $\bm{k}$-space (with doubled unit cell), the bilinear part reads
\begin{align*}
  H_{\textrm{bl}} &= 
  \frac{3}{2}\bigl(J_{1,\perp}+J_{5,\perp}\bigr)\bigl[\tilde{S}_{\textrm{A}}^+(\bm{k}=0) \tilde{S}_{\textrm{B}}^-(\bm{k}=0)+\tilde{S}_{\textrm{C}}^+(\bm{k}=0) \tilde{S}_{\textrm{D}}^-(\bm{k}=0)r\bigr]
  \\&\quad+3J_{4,\perp}\bigl[\tilde{S}_{\textrm{A}}^+(\bm{k}=0) \tilde{S}_{\textrm{C}}^-(\bm{k}=0)+\tilde{S}_{\textrm{B}}^+(\bm{k}=0) \tilde{S}_{\textrm{D}}^-(\bm{k}=0)\bigr]\\
  &\quad+\frac{1}{2}\bigl(J_{2,\perp}+3J_{6,\perp}\bigr)\bigl[\tilde{S}_{\textrm{A}}^+(\bm{k}=0) \tilde{S}_{\textrm{D}}^-(\bm{k}=0)+\tilde{S}_{\textrm{B}}^+(\bm{k}=0) \tilde{S}_{\textrm{C}}^-(\bm{k}=0)\bigr]\\
  &\quad+\textrm{h.c}+\cdots,
\end{align*}
where $\cdots$ refers to other terms irrelevant to determining the classical ground state below.

For the in-plane order 
\begin{align}
  \langle \tilde{S}_{\bm{r}}^z\rangle&=0,\\
  \langle \tilde{S}_{\bm{r}}^+\rangle&=\tilde{S}\e^{\ii\phi_{\mu_{\bm{r}}}},
\end{align}
where $\mu_{\bm{r}}=\A,\B,\C,\D$ is the sublattice index of $\bm{r}$, we need to minimize
\begin{align*}
  f(\phi_{\A},\phi_{\B},\phi_{\C},\phi_{\D})=\;
  &3\tilde{S}^2\bigl(J_{1,\perp}+J_{5,\perp}\bigr)\bigl\{\cos(\phi_{\A}-\phi_{\B})+\cos(\phi_{\C}-\phi_{\D})\bigr\}\\
  &6\tilde{S}^2J_{4,\perp}\bigl\{\cos(\phi_{\A}-\phi_{\C})+\cos(\phi_{\B}-\phi_{\D})\bigr\}\\
  &\tilde{S}^2\bigl(J_{2,\perp}+3J_{6,\perp}\bigr)\bigl\{\cos(\phi_{\A}-\phi_{\D})+\cos(\phi_{\B}-\phi_{\C})\bigr\}\\
  &+\alpha_{6} \tilde{S}^6\bigl[\cos(3\phi_{\A}+3\phi_{\B}-\phi_{6})+\cos(3\phi_{\C}+3\phi_{\D}-\phi_{6})\bigr]\\
  &+ g\mu_{\textrm{B}}B\tilde{S}\sum_{\mu}\cos(\phi_{\mu}-\phi_{b})
\end{align*}
Using the parameters (all in meV)
\begin{equation}
  J_{1,\perp}=-6.36,\quad J_{2,\perp}=-0.33,\quad J_{3,\perp}=0.78,\quad J_{4,\perp}=0.11,\quad J_{5,\perp}=-0.39,\quad J_{6,\perp}=0.79
  \label{eqnparam}
\end{equation}
we have $\phi_{\A}=\phi_{\B}$ and $\phi_{\C}=\phi_{\D}$. Additionally,
\begin{itemize}
  \item $H_{\textrm{bl}}$ is minimized by $\phi_{\A}-\phi_{\C}=\text{\textpi}$
%  \item $H_{4}$ is irrelevant for an in-plane order
  \item $H_{6}$ is minimized by (with $\tilde{\alpha}_{6}>0$)
    \begin{align*}
      \phi_{\A}-\phi_{\C}&=\text{\textpi}/3+2\text{\textpi} n/3\\
      \phi_{\A}+\phi_{\C} &= \phi_{6}/3+2\text{\textpi} m/3,\quad n,m\in\mathbb{Z}
    \end{align*}
\end{itemize}
Therefore, in the absence of magnetic fields, $H_{\textrm{bl}}+H_{6}$ yields
\begin{align}
  \phi_{\A}&=\phi_{\B}= \frac{\phi_{6}}{6}+\frac{\text{\textpi}}{2}+\frac{m\text{\textpi}}{3}\\
  \phi_{\C} &= \phi_{\D}=\frac{\phi_{6}}{6}-\frac{\text{\textpi}}{2}+\frac{m\text{\textpi}}{3}
\end{align}
\noindent Note that $\phi_{\A,\B}$ is always antiparallel to $\phi_{\C,\D}$ as experimentally determined.

\begin{figure}[b]
  \centering
  \includegraphics[width=0.7\textwidth]{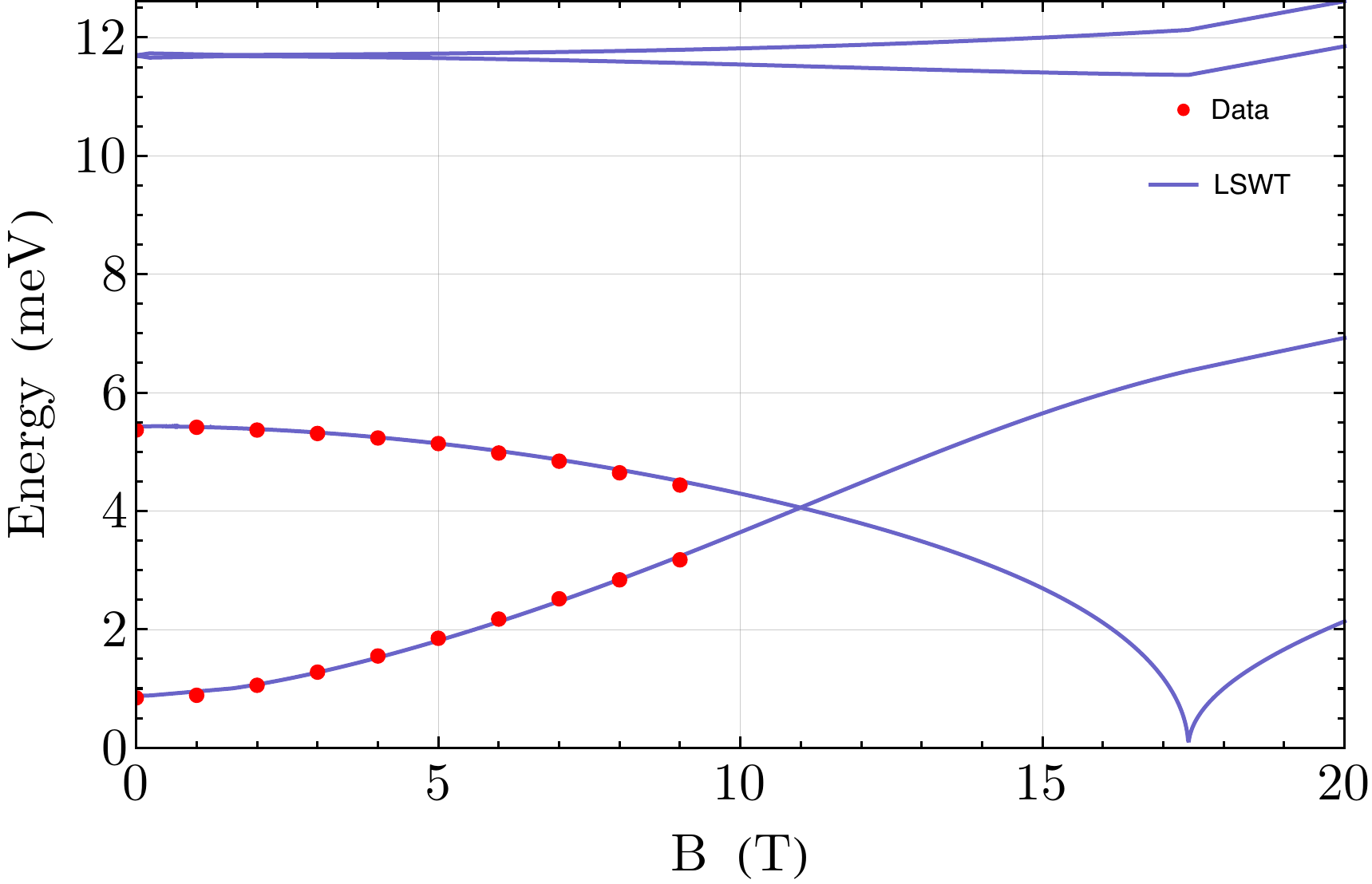}
  \caption{Linear spin wave theory of the $\tilde{S}=1/2$ calculation of the magnon spectrum using the parameters in equation \ref{eqnparam} in addition to $g=2.73$, $\phi_b=\text{\textpi}/2$, $\alpha_{6}=\SI{46}{\micro\electronvolt}$ and $\phi_{6}=\text{\textpi}$.}
  \label{fig:s1/2fielddep}
\end{figure}

\begin{figure}[t]
  \centering
  \includegraphics[width=0.7\textwidth]{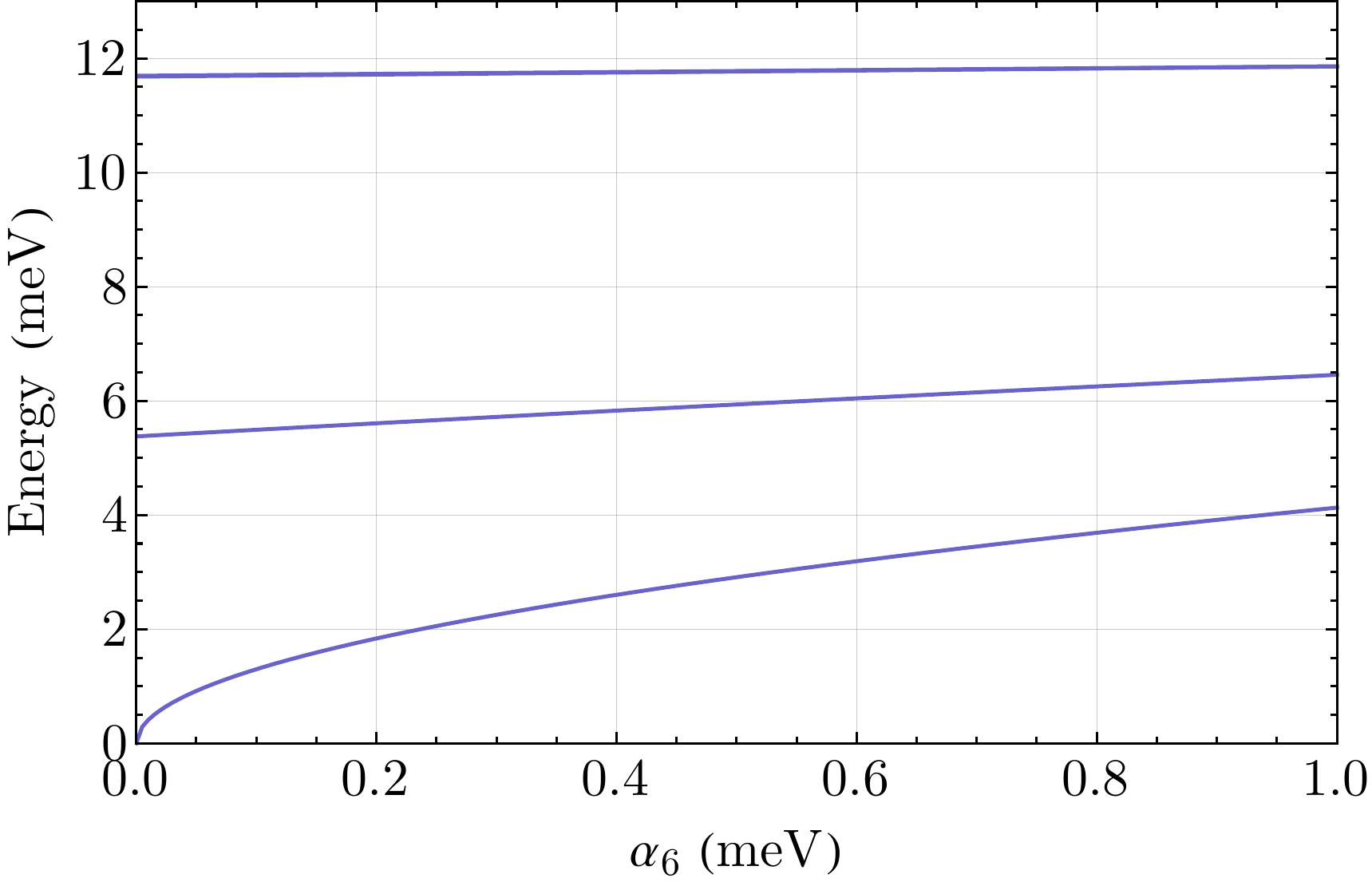}
  \caption{Effect of the ring-exchange term at $B=0$ on the zone-center magnon frequencies using the $\tilde{S}=1/2$ model. The moments remain in-plane, are ferromagnetic within each layer, and antiferromagnetic between layers for this range of values of $\alpha_6$.}
  \label{fig:ringdep}
\end{figure}

\subsection*{Effects of ring-exchange on LSWT}
Below, $H_{6}$ is expanded around an in-plane order $\phi_{\A,\B}=0$ and $\phi_{\C,\D}=\text{\textpi}$ to obtain the quadratic terms. The ring term can be expanded to obtain: 
\begin{align*}
  H_{6} &= \tilde{\alpha}_{6}\e^{-\ii\tilde{\phi}_{6}}\sum_{\bm{r}}\prod_{i=1}^{6}S_{\bm{r}+\bm{\delta}_{i}^{\textrm{ring}}}^{+} + \textrm{h.c.}\\
   &= \begin{multlined}[t]
     -\tilde{\alpha}_{6}\tilde{S}^{4}\e^{-\ii\tilde{\phi}_{6}}\e^{\ii(3\phi_{\A}+3\phi_{\B})}\\
     \times\sum_{\bm{r}\in \A\B\textrm{ layer}}\biggl[\frac{1}{2}\sum_{i=1}^{6}\Bigl[{\bigl(\tilde{S}_{\bm{r}+\bm{\delta}^{\textrm{ring}}_{i}}^{y'}\bigr)}^2+{\bigl(\tilde{S}_{\bm{r}+\bm{\delta}^{\textrm{ring}}_{i}}^{z}\bigr)}^2\Bigr]+\frac{1}{2}\sum_{i\ne j}^{6}\tilde{S}_{\bm{r}+\bm{\delta}^{\textrm{ring}}_{i}}^{y'}\tilde{S}_{\bm{r}+\bm{\delta}^{\textrm{ring}}_{j}}^{y'}\biggr]+\textrm{h.c.}
     \end{multlined}\\
     &+[\A,\B\rightarrow\C,\D]
  \end{align*}

The results of fitting the magnetic field dependence using the $\tilde{S}=1/2$ model described here are shown in figure \ref{fig:s1/2fielddep} using the parameters listed in eqn. \ref{eqnparam} and in Table 1 of the main text. In addition, the dependence of the four zone-center magnon energies at zero magnetic field on the strength of the ring-exchange parameter $\alpha_6$ are shown in figure \ref{fig:ringdep}. Note that the energy of the pseudo-Goldstone mode seems to follow a $\sim\sqrt{\tilde{\alpha}_6}$ dependence on the ring-exchange strength for small values of $\alpha_6$. This is why such a small value of the $\alpha_6$ parameter can open such a sizable gap.

\subsection*{Magnon normal modes}
As shown in equation \ref{magnon wf:goldstone} without the ring-exchange interaction the Goldstone mode only has dynamical components in the honeycomb plane. We write the normal modes in the form $\tilde{S}\left[(\tilde{S}^y_{1,1},\tilde{S}^z_{1,1}),(\tilde{S}^y_{1,2},\tilde{S}^z_{1,2}),(\tilde{S}^y_{2,1},\tilde{S}^z_{2,1}),(\tilde{S}^y_{2,2},\tilde{S}^z_{2,2})\right]$. Here $\tilde{S}^\beta_{\mu,\nu}$ is the dynamical spin component along the $\beta$-axis, layer $\mu$ and sublattice $\nu$. Including the ring-exchange interaction the normal modes have the following form:

\begin{align*}
    &\SI{0.9}{meV} &\rightarrow &\quad\quad  \tilde{S}\left[(0.5,0.03\ii),(0.5,0.03\ii),(-0.5,0.03\ii),(-0.5,0.03\ii)\right]\\
    &\SI{5.3}{meV} &\rightarrow &\quad\quad  \tilde{S}\left[(0.45,0.23\ii),(0.45,0.23\ii),(0.45,-0.23\ii),(0.45,-0.23\ii)\right]\\
    &\SI{11.6}{meV} &\rightarrow &\quad\quad   \tilde{S}\left[(0.23\ii,-0.45),(-0.23\ii,0.45),(-0.23\ii,-0.45),(0.23\ii,0.45)\right]\\
    &\SI{11.7}{meV} &\rightarrow &\quad\quad  \tilde{S}\left[(-0.24\ii,0.43),(0.24\ii,-0.43),(-0.24\ii,-0.43),(0.24\ii,0.43)\right]
\end{align*}
%Note that the net magnetic dipole moments of the two lowest energy magnons are non-zero, $\sum_{\mu,\nu}^z\neq 0$ and $\sum_{\mu,\nu}^y\neq 0$, respectively, while it is zero for the two highest energy modes because $\sum_{\mu,\nu}^{y,z}= 0$. 
These explain the direction of the polarization of the THz magnetic field that excite these modes. The lowest, pseudo-Goldstone, magnon is only excited with $\bm{h}_\omega\parallel [001]$ because $\sum_{\mu,\nu}\tilde{S}_{\mu\nu}^z\neq 0$ in this mode. Because the second mode has $\sum_{\mu,\nu}\tilde{S}_{\mu\nu}^y\neq 0$ then it can be excited by $\bm{h}_\omega \perp [001]$, as observed experimentally. The normal modes also tell us that the two highest energy magnons can not be observed in THz experiments by the magnetic field of the THz because they have zero net magnetic moment $\sum_{\mu,\nu}\tilde{S}_{\mu\nu}^{y,z}= 0$. We note too that these four modes all have elliptical magnetic moment oscillations as the $y$ and $z$ components are always $\text{\textpi}/2$ out of phase.

\section{Supplementary Note 5: flavor Wave model}
\begin{figure}[htpb]
  \centering
  \includegraphics[width=0.4\textwidth]{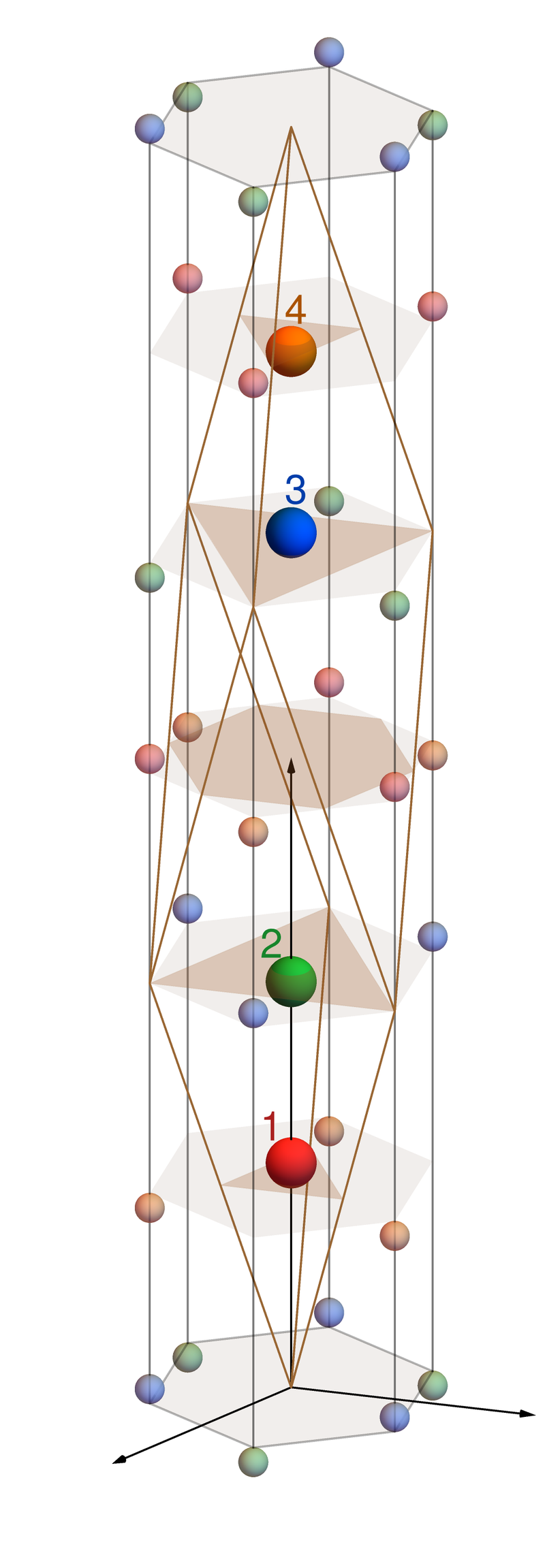}
  \caption{Magnetic unit cell (brown) used in the flavor wave theory calculation and the sublattice definitions. There are four Co atoms per unit cell, labeled 1 through 4, and color coded by sublattice.}
  \label{fig:unitcell}
\end{figure}

Here we show the details of the flavor wave model. First we set a Cartesian coordinates $Oxyz$ as $\hat{\bm{x}}=\hat{\bm{b}}$ and $\hat{\bm{z}}=\hat{\bm{c}}$, where $\bm{a}$, $\bm{b}$, and $\bm{c}$ are the hexagonal lattice vectors in obverse setting.  We will first discuss the single ion physics. The Co$^{2+}$ ion has a [Ar]$3\mathrm{d}^7$ electron configuration. An isolated Co$^{2+}$ has the ground state $^4$F$_{9/2}$ of the $^4$F term, which implies that $L=3$ and $S=3/2$. The next lowest term is $^4$P with the lowest state $^4$P$_{5/2}$ which is $\SI{15202.6(2)}{\centi\meter ^{-1}}$ ($\SI{1.88488(3)}{\electronvolt}$) above the ground state\cite{CoSpectra}. In a crystal field, the ground state mostly comes from the $^4$F states. An experiment shows that $^4$P contributes $6\%$ of the actual ground state in CoO\cite{IRCoO}, but we will ignore the contribution from this term in CoTiO$_3$. An octahedral crystal field environment, where the 3-fold axis is along $z$, produces the following interaction:$$
H_{\mathrm{cf}}=B_{\mathrm{cf}}\left(\mathcal{O}_4^0-20\sqrt{2}\mathcal{O}_4^3\right)
$$
where $\mathcal{O}_4^0=35L_z^4-30L(L+1)L_z^2+25L_z^2-6L(L+1)+3L^2(L+1)^2$ \\ and $\mathcal{O}_4^3=\frac{1}{4}\left(L_z\left(L_+^3+L_-^3\right)+\left(L_+^3+L_-^3\right)L_z\right)$\cite{AandBbook}. The crystal field splits the $2L+1=7$ degenerate orbitals into a lowest triplet $\Gamma_4$, which will be expressed as an effective $l=1$, an excited triplet $\Gamma_5$ at $720B_{\mathrm{cf}}$ and a singlet $\Gamma_2$ at $1620B_{\mathrm{cf}}$. The wavefunctions of the $\Gamma_4$ ground triplet are$$
\begin{aligned}
    &\left|l_z=1\right>_{\Gamma _4}&=&\sqrt{\frac{1}{6}}\left|L_z=1\right>-\sqrt{\frac{5}{6}}\left|L_z=-2\right>\\
    &\left|l_z=0\right>_{\Gamma _4}&=&\sqrt{\frac{5}{18}}\left|L_z=-3\right>-\frac{2}{3}\left|L_z=0\right>-\sqrt{\frac{5}{18}}\left|L_z=3\right>\\
    &\left|l_z=-1\right>_{\Gamma _4}&=&\sqrt{\frac{1}{6}}\left|L_z=-1\right>+\sqrt{\frac{5}{6}}\left|L_z=2\right>
\end{aligned}
$$
with a relation of $\left<l_z\right|L_z\left|l_z\right>=-3l_z/2$ \cite{AandBbook}. Since we have the relation that Coulomb interaction $>$ cubic crystal field $\gg$ spin-orbital coupling $\approx$ trigonal distortion of crystal field $\gtrsim$ interaction between ions, we only consider a space of $(2l+1)(2S+1)=12$ dimension for each ion. We choose the unit cell to be the rhombohedral unit cell with sublattice index $i$ defined as in figure \ref{fig:unitcell}. We work in a space that is the product of the local Hilbert spaces of each individual atom $(\bm{r},i)$, where $\bm{r}=a_1(\bm{a}/3+2\bm{b}/3+2\bm{c}/3)+a_2(-2\bm{a}/3-\bm{b}/3+2\bm{c}/3)+a_3(\bm{a}/3-\bm{b}/3+2\bm{c}/3)$, $a_{\alpha}\in\mathbb{Z}(\alpha=1,2,3)$, is the position of the unit cell and $i$ is the sublattice index.
The Hamiltonian is $$
H=H_0+H_{\mathrm{Z}}+H_{\mathrm{bl}}+H_{\mathrm{ring}}+H_{\mathrm{bq}}
$$
where $H_0$ describes the trigonal distortion ($\delta$) and spin-orbital coupling ($\lambda$)
$$H_0=\sum_{\bm{r},i}(3\lambda/2) \bm{S}_{\bm{r},i}\cdot\bm{l}_{\bm{r},i}+\delta \left(\left(l^z_{\bm{r},i}\right) ^2 -\frac{2}{3}\right)
$$
which sums over the $i$-th atom($i=1,2,3,4$) at unit cell position $\bm{r}$.
$H_{\mathrm{Z}}$ describes the interaction with external magnetic field $\bm{B}$:
$$H_{\mathrm{Z}}=\mu_\mathrm{B}\sum_{\bm{r},i}\bm{B}\cdot\left(2\bm{S}_{\bm{r},i}-\tfrac{3}{2}\bm{l}_{\bm{r},i}\right)
$$
where $\mu_\mathrm{B}=\SI{0.05788}{\milli\electronvolt/\tesla}$ is the Bohr magneton. $H_{\mathrm{bl}}$ describes the interaction between the atomic magnetic moments. We write down below expressions for different types of interaction that contribute to $H_{\mathrm{bl}}$.
For the $J_1$ Heisenberg interaction (see figure 1 of main text), we have the Hamiltonian to be 
$$
\begin{aligned}
H_{J_1}=&\frac{J_1}{2}\sum_{\bm{r}}\left(\bm{S}_{\bm{r},1}\cdot \bm{S}_{\bm{r}+\frac{\bm{a}}{3}+\frac{2\bm{b}}{3}-\frac{4\bm{c}}{3},4}+\bm{S}_{\bm{r},1}\cdot \bm{S}_{\bm{r}+\frac{\bm{a}}{3}-\frac{\bm{b}}{3}-\frac{4\bm{c}}{3},4}+\bm{S}_{\bm{r},1}\cdot \bm{S}_{\bm{r}-\frac{2\bm{a}}{3}-\frac{\bm{b}}{3}-\frac{4\bm{c}}{3},4}\right)+\\
&\frac{J_1}{2}\sum_{\bm{r}}\left(\bm{S}_{\bm{r},2}\cdot \bm{S}_{\bm{r}-\frac{\bm{a}}{3}-\frac{2\bm{b}}{3}-\frac{2\bm{c}}{3},3}+\bm{S}_{\bm{r},2}\cdot \bm{S}_{\bm{r}-\frac{\bm{a}}{3}+\frac{\bm{b}}{3}-\frac{2\bm{c}}{3},3}+\bm{S}_{\bm{r},2}\cdot \bm{S}_{\bm{r}+\frac{2\bm{a}}{3}+\frac{\bm{b}}{3}-\frac{2\bm{c}}{3},3}\right)+\\
&\frac{J_1}{2}\sum_{\bm{r}}\left(\bm{S}_{\bm{r},3}\cdot \bm{S}_{\bm{r}+\frac{\bm{a}}{3}+\frac{2\bm{b}}{3}+\frac{2\bm{c}}{3},2}+\bm{S}_{\bm{r},3}\cdot \bm{S}_{\bm{r}+\frac{\bm{a}}{3}-\frac{\bm{b}}{3}+\frac{2\bm{c}}{3},2}+\bm{S}_{\bm{r},3}\cdot \bm{S}_{\bm{r}-\frac{2\bm{a}}{3}-\frac{\bm{b}}{3}+\frac{2\bm{c}}{3},2}\right)+\\
&\frac{J_1}{2}\sum_{\bm{r}}\left(\bm{S}_{\bm{r},4}\cdot \bm{S}_{\bm{r}-\frac{\bm{a}}{3}-\frac{2\bm{b}}{3}+\frac{4\bm{c}}{3},1}+\bm{S}_{\bm{r},4}\cdot \bm{S}_{\bm{r}-\frac{\bm{a}}{3}+\frac{\bm{b}}{3}+\frac{4\bm{c}}{3},1}+\bm{S}_{\bm{r},4}\cdot \bm{S}_{\bm{r}+\frac{2\bm{a}}{3}+\frac{\bm{b}}{3}+\frac{\bm{c}}{3},1}\right)
\end{aligned}
$$
Notice that each bond is counted twice with strength $\frac{J_1}{2}$. For $J_4$ Heisenberg interaction, we will write down

\begin{alignat*}{2}
H_{J_4}=& & \frac{J_4}{2}\sum_{\bm{r}}&\left(\bm{S}_{\bm{r},1}\cdot \bm{S}_{\bm{r}+\frac{\bm{a}}{3}+\frac{2\bm{b}}{3}-\frac{4\bm{c}}{3},3}+\bm{S}_{\bm{r},1}\cdot \bm{S}_{\bm{r}+\frac{\bm{a}}{3}-\frac{\bm{b}}{3}-\frac{4\bm{c}}{3},3}+\bm{S}_{\bm{r},1}\cdot \bm{S}_{\bm{r}-\frac{2\bm{a}}{3}-\frac{\bm{b}}{3}-\frac{4\bm{c}}{3},3}+\right.\\
&&{} &\left.\bm{S}_{\bm{r},1}\cdot \bm{S}_{\bm{r}-\frac{\bm{a}}{3}-\frac{2\bm{b}}{3}-\frac{2\bm{c}}{3},3}+\bm{S}_{\bm{r},1}\cdot \bm{S}_{\bm{r}-\frac{\bm{a}}{3}+\frac{\bm{b}}{3}-\frac{2\bm{c}}{3},3}+\bm{S}_{\bm{r},1}\cdot \bm{S}_{\bm{r}+\frac{2\bm{a}}{3}+\frac{\bm{b}}{3}-\frac{2\bm{c}}{3},3}\right)\\
&&{} &+\left[\bm{S}_{\bm{r},4}\leftrightarrow \bm{S}_{\cdot,1}\right]+\left[\bm{S}_{\bm{r},2}\leftrightarrow \bm{S}_{\cdot,3}\right]+\left[\bm{S}_{\bm{r},3}\leftrightarrow \bm{S}_{\cdot,2}\right]
\end{alignat*}

Here in $\bm{S}_{\cdot,\#}$, $\cdot$ stands for one of the six out of plane next nearest neighbor vectors, and $\#$ is one of the four magnetic sublattices. Each bond is also counted twice with strength $\frac{J_4}{2}$. For the ring exchange, we will write down

\begin{alignat*}{2}
H_{\mathrm{ring}}=&\sum_{\bm{r}}&\left( rS^+_{\bm{r}+\frac{\bm{a}}{3}+\frac{2\bm{b}}{3}+\frac{2\bm{c}}{3},1}  S^+_{\bm{r}-\frac{\bm{a}}{3}+\frac{\bm{b}}{3}-\frac{2\bm{c}}{3},4} S^+_{\bm{r}-\frac{\bm{a}}{3}-\frac{\bm{b}}{3}+\frac{2\bm{c}}{3},1} S^+_{\bm{r}-\frac{\bm{a}}{3}-\frac{2\bm{b}}{3}-\frac{2\bm{c}}{3},4} S^+_{\bm{r}+\frac{\bm{a}}{3}-\frac{\bm{b}}{3}+\frac{2\bm{c}}{3},1} S^+_{\bm{r}+\frac{2\bm{a}}{3}+\frac{\bm{b}}{3}-\frac{2\bm{c}}{3},4}+ \right. &\\
&{}& \left. r^*S^-_{\bm{r}+\frac{\bm{a}}{3}+\frac{2\bm{b}}{3}+\frac{2\bm{c}}{3},1}  S^-_{\bm{r}-\frac{\bm{a}}{3}+\frac{\bm{b}}{3}-\frac{2\bm{c}}{3},4} S^-_{\bm{r}-\frac{\bm{a}}{3}-\frac{\bm{b}}{3}+\frac{2\bm{c}}{3},1} S^-_{\bm{r}-\frac{\bm{a}}{3}-\frac{2\bm{b}}{3}-\frac{2\bm{c}}{3},4} S^-_{\bm{r}+\frac{\bm{a}}{3}-\frac{\bm{b}}{3}+\frac{2\bm{c}}{3},1} S^-_{\bm{r}+\frac{2\bm{a}}{3}+\frac{\bm{b}}{3}-\frac{2\bm{c}}{3},4}  \right)&+\\
&\sum_{\bm{r}}&\left( rS^+_{\bm{r}+\frac{\bm{a}}{3}+\frac{2\bm{b}}{3}+\frac{2\bm{c}}{3},3}  S^+_{\bm{r}-\frac{\bm{a}}{3}+\frac{\bm{b}}{3}+\frac{4\bm{c}}{3},2} S^+_{\bm{r}-\frac{\bm{a}}{3}-\frac{\bm{b}}{3}+\frac{2\bm{c}}{3},3} S^+_{\bm{r}-\frac{\bm{a}}{3}-\frac{2\bm{b}}{3}+\frac{4\bm{c}}{3},2} S^+_{\bm{r}+\frac{\bm{a}}{3}-\frac{\bm{b}}{3}+\frac{2\bm{c}}{3},3} S^+_{\bm{r}+\frac{2\bm{a}}{3}+\frac{\bm{b}}{3}+\frac{4\bm{c}}{3},2}+ \right. &\\
&{}& \left. r^*S^-_{\bm{r}+\frac{\bm{a}}{3}+\frac{2\bm{b}}{3}+\frac{2\bm{c}}{3},3}  S^-_{\bm{r}-\frac{\bm{a}}{3}+\frac{\bm{b}}{3}+\frac{4\bm{c}}{3},2} S^-_{\bm{r}-\frac{\bm{a}}{3}-\frac{\bm{b}}{3}+\frac{2\bm{c}}{3},3} S^-_{\bm{r}-\frac{\bm{a}}{3}-\frac{2\bm{b}}{3}+\frac{4\bm{c}}{3},2} S^-_{\bm{r}+\frac{\bm{a}}{3}-\frac{\bm{b}}{3}+\frac{2\bm{c}}{3},3} S^-_{\bm{r}+\frac{2\bm{a}}{3}+\frac{\bm{b}}{3}+\frac{4\bm{c}}{3},2}  \right)&
\end{alignat*}
Notice that each ring is counted once with two conjugate terms.

To obtain the ground state at zero temperature, we apply the mean field approximation, which means the wave function $\left|\Psi\right>$ can be written as a product
$$\left|\Psi\right>=\prod_{\bm{r},i}\left|\psi_{\bm{r},i}\right>
$$
and we apply the long-range order approximation so that
$$\left|\psi_i\right>=\left|\psi_{\bm{r},i}\right>,\quad \forall \bm{r}
$$
Now the wave functions must satisfy $$
h_i\left|\psi_i\right>=E_i\left| \psi_i\right>
$$
where $h_i$ is the mean field Hamiltonian and $E_i$ is the smallest eigenvalue of $h_i$. We can get the expression for $h_i$ is$$
h_i=h_0+h_{i,\mathrm{Z}}+h_{i,J_1}+h_{i,J_4}+h_{i,\mathrm{6}}+h_{i,\mathrm{bq}}
$$
$$h_0=\frac{3\lambda}{2}\left(S_xl_x+S_yl_y+S_zl_z\right)-\delta\left(l_z^2-\frac{2}{3}\right)
$$
$$h_{i,\mathrm{Z}}=\mu_\mathrm{B}\left(B_x\left(2S_x-\frac{3}{2}l_x\right)+B_y\left(2S_y-\frac{3}{2}l_y\right)+B_z\left(2S_z-\frac{3}{2}l_z\right)\right)
$$
And for example, the 1-st Co will have:
$$h_{1,J_1}=3J_1 \left(S_x \left<\psi_4\right|S_x\left|\psi_4\right>+S_y \left<\psi_4\right|S_y\left|\psi_4\right>+S_z \left<\psi_4\right|S_z\left|\psi_4\right>\right)
$$
$$h_{1,J_4}=6J_4 \left(S_x \left<\psi_3\right|S_x\left|\psi_3\right>+S_y \left<\psi_3\right|S_y\left|\psi_3\right>+S_z \left<\psi_3\right|S_z\left|\psi_3\right>\right)
$$
$$h_{1,\mathrm{ring}}=3r S^+\left(\left<\psi_4\right|S_+\left|\psi_4\right>\right)^3\left(\left<\psi_1\right|S_+\left|\psi_1\right>\right)^2+3r^* S^-\left(\left<\psi_4\right|S_-\left|\psi_4\right>\right)^3\left(\left<\psi_1\right|S_-\left|\psi_1\right>\right)^2
$$

Note that, all matrices are 12 dimensional. For example, in the second equation, $S_zl_z$ should be interpreted as $(\mathrm{diag}(3/2,1/2,-1/2,-3/2)\otimes\mathrm{diag}(1,1,1))\cdot(\mathrm{diag}(1,1,1,1)\otimes\mathrm{diag}(1,0,-1))$. To get the wave function at $\SI{0}{\kelvin}$ and a certain magnetic field, we use the imaginary time propagation algorithm. We start from a state $\mathbf{x}_0=\left(\left|\psi_1\right>,\left|\psi_2\right>,\left|\psi_3\right>,\left|\psi_4\right>\right)$, and do:\\
\indent\indent(1)for state $\mathbf{x}_n$, calculate mean field Hamiltonian $\mathbf{h}_n=\left(h_1\left(\mathbf{x}_n\right),h_2\left(\mathbf{x}_n\right),h_3\left(\mathbf{x}_n\right),h_4\left(\mathbf{x}_n\right)\right)$\\
\indent\indent(2)check if $\forall i$, $\mathbf{x}_{n+1,i}$ is close to the eigenstate of $\mathbf{h}_{n,i}$ which corresponds to the smallest eigenvalue,\\
\indent \indent  \indent if so, stop, $\mathbf{x}_n$ is the mean field ground state,\\
\indent  \indent \indent if not, go to step (3)\\
\indent\indent(3)time-evolve: $\mathbf{y}_{n+1}=((1,1,1,1)-\varepsilon_{n}\mathbf{h}_n)\cdot\mathbf{x}_n$, where $\varepsilon_{n}$ is a small positive number\\
\indent\indent(4)Normalize the wave function, the $i$-th component of $\mathbf{x}_{n+1,i}=\frac{\mathbf{y}_{n+1,i}}{\sqrt{|\mathbf{y}_{n+1,i}|^2}}$, \\
\indent go back to step(1).\\

\begin{figure}[htpb]
  \centering
  \includegraphics[width=0.8\textwidth]{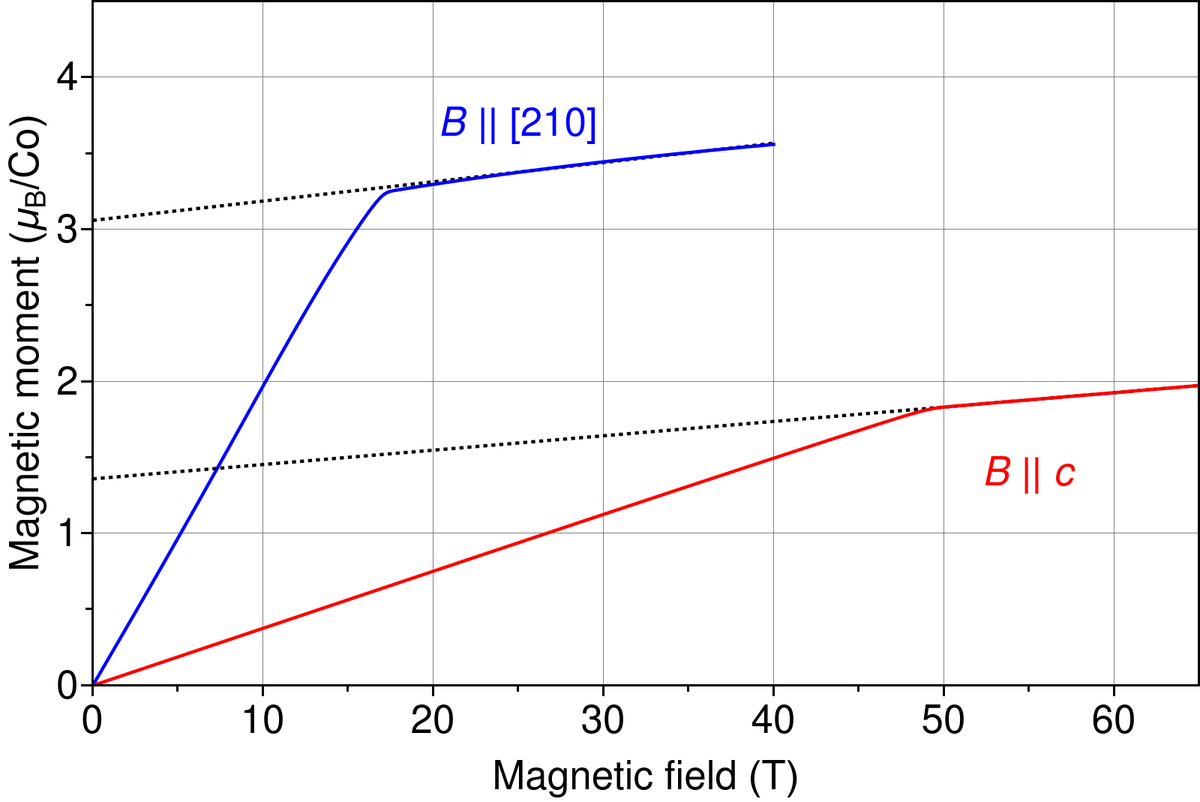}
  \caption{Calculated magnetization, within the flavor wave model, at $\SI{0}{\kelvin}$. The ground state at $\SI{0}{T}$ is assumed to be with the magnetic moment along $\hat{b}$ as a single domain. 
  Dotted lines are linear fit of $B> \SI{22}{T}$ for in-plane field and $B> \SI{50}{T}$ region for the out-of-plane field. This calculation should be compared to Fig.5(a) of ref.\cite{permeability2021}.}
  \label{fig:magneticmement}
\end{figure}

Using this algorithm, we can find the mean-field ground state. Then we calculate the expectation $\mu_\mathrm{B}\left<\psi_i\left|2\bm{S}-3\bm{l}/2\right|\psi_i\right>$ to get the magnetic moment of each ion. After adding the magnetic moment of all four atoms, one can get the net magnetic dipole of a unit cell. The calculated results are shown in figure \ref{fig:magneticmement} for two directions of magnetic field. This should be compared to experimental results at Fig.5(a) of ref.\cite{permeability2021}.

\begin{figure}[b]
  \centering
  \includegraphics[width=0.7\textwidth]{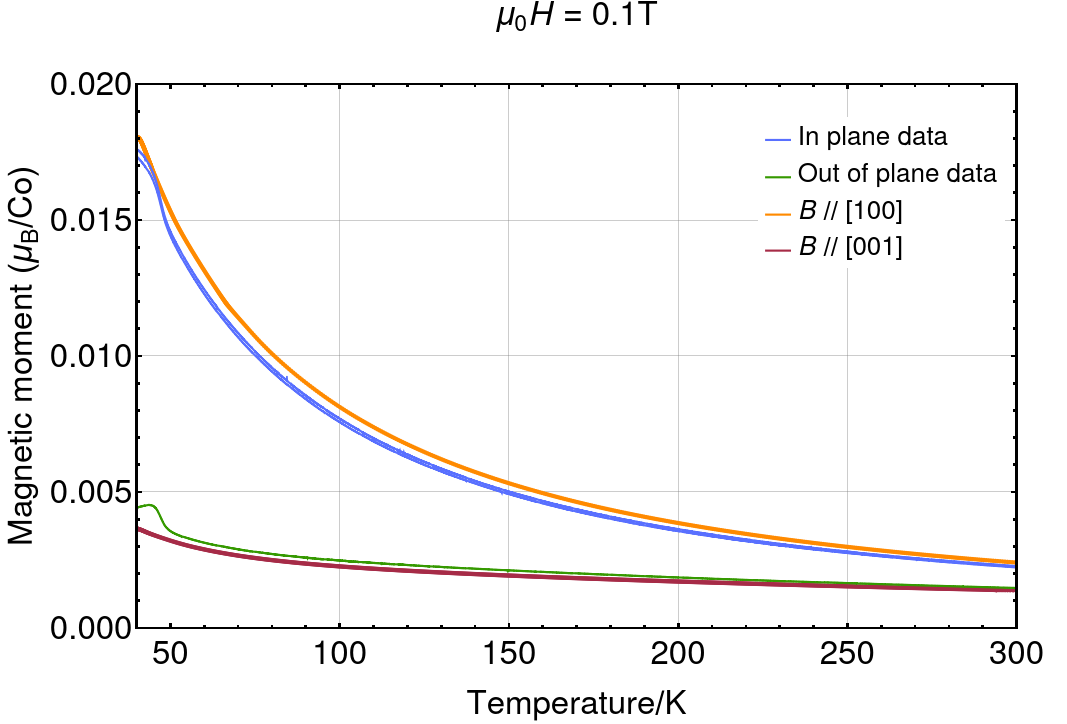}
  \caption{Comparison between experiment and flavor wave model calculation of the temperature dependence of the magnetization above $T_{\mathrm{N}}$. Blue and green lines are from experiment, orange and brown are from calculations.}
  \label{fig:magnetization}
\end{figure}

To get the temperature dependence, we only consider the two-spin correlation function and use density matrix theory. The state is described by $4+\dbinom{4}{2}=10$ density matrices. For example, to get the correlation between ion 1 and ion 4, we first write down the Hamiltonian for Co 1 and Co 4:
$$h_{14}=h_{14,0}+h_{14,\mathrm{Z}}+h_{14,J_1}+h_{14,J_4}+h_{14,\mathrm{ring}}
$$
$$h_{14,0}=\left(\frac{3\lambda}{2}\bm{S}\cdot\bm{l}-\delta\left(l_z^2-\frac{2}{3}\right) \right)\otimes I+I\otimes\left(\frac{3\lambda}{2}\bm{S}\cdot\bm{l}-\delta\left(l_z^2-\frac{2}{3}\right) \right)
$$
$$h_{14,\mathrm{Z}}=\left(\mu_\mathrm{B}\bm{B}\cdot\left(2\bm{S}-\frac{3}{2}\bm{l}\right)\right)\otimes I+I\otimes\left(\mu_\mathrm{B}\bm{B}\cdot\left(2\bm{S}-\frac{3}{2}\bm{l}\right)\right)
$$
$$h_{14,J_1}=3J_1\left(\left(S_x\otimes I\right)\left(I\otimes S_x\right)+\left(S_y\otimes I\right)\left(I\otimes S_y\right)+\left(S_z\otimes I\right)\left(I\otimes S_z\right)\right)
$$
$$h_{14,J_4}=6J_4 \left( \left<\psi_3\right|\bm{S}\left|\psi_3\right>\cdot\bm{S}\right)\otimes I+12J_4 I \otimes \left( \left<\psi_2\right|\bm{S}\left|\psi_2\right>\cdot\bm{S}\right)
$$
$$
\begin{aligned}
h_{14,\mathrm{ring}}=& 3\times3r(S_+\otimes I )(I \otimes S_+)\left(\left<\psi_1\right|S_+\left|\psi_1\right>\right)^2\left(\left<\psi_4\right|S_+\left|\psi_4\right>\right)^2\\
&+9r^*(S_+\otimes I )(I \otimes S_-)\left(\left<\psi_1\right|S_-\left|\psi_1\right>\right)^2\left(\left<\psi_4\right|S_-\left|\psi_4\right>\right)^2
\end{aligned}
$$

\noindent where $I$ is a $12\times12$ identity matrix, $\otimes$ is the Kronecker product. $S_z = \mathrm{diag}(3/2, 1/2, -1/2, -3/2)\otimes \mathrm{diag}(1, 1, 1)$, $l_z=\mathrm{diag}(1, 1, 1, 1)\otimes \mathrm{diag}(1, 0, -1)$. After getting the joint Hamiltonian $h_{14}$, the density matrix will be $$
\rho_{14}=\frac{\e^{-\beta h_{14}}}{\mathrm{tr}\left( \e^{-\beta h_{14}}\right)}
$$
where $\beta =1/(k_{\mathrm{B}} T)$, $T$ is the temperature and $k_{\mathrm{B}}=\SI{0.08617}{\milli\electronvolt/\kelvin}$ is the Boltzmann constant. After tracing out the 4-th ion, we can get the density matrix for 1-st ion: $\rho _1 = \mathrm{tr}_4 (\rho _{14})$. The total magnetic moment of one unit cell will be $$
\bm{m}=\mu_{\mathrm{B}}\sum_{i=1}^4 \mathrm{tr}\left(\rho_i \left(2\bm{S}-3\bm{l}/2\right)\right)
$$
Now we can calculate the temperature dependence. This is shown in figure \ref{fig:magnetization} for two directions of the applied magnetic together with the experimental data for temperatures higher than 40 K.

Using the ground state at $\SI{0}{\kelvin}$, we can solve for the magnon dispersion. For a ground state $\mathbf{x}=\left(\left|\psi_1\right>,\left|\psi_2\right>,\left|\psi_3\right>,\left|\psi_4\right>\right)$,  we first calculate the mean-field Hamiltonian $\mathbf{h}_i$, and diagonalize it to get $$
\mathbf{h}_i=\left|\psi_i\right>E_{i,0}\left<\psi_i\right|+\sum_{m=1}^{d-1}\left|\psi_{i,m}\right>E_{i,m}\left<\psi_{i,m}\right|
$$
where $d=(2l+1)(2S+1)=12$ is the dimension of the Hilbert space of each atom. We also define $\left|\psi_{i,0}\right>=\left|\psi_{i}\right>$. Now for any operator $\hat{X}$ acting on atom $(\bm{r},i)$ we can have a bosonic operator expansion \cite{generalizedSWT}
$$
\hat{X}\approx X_{\bm{r}i,00}+\sum_{m=1}^{d-1}\left(\hat{b}_{\bm{r},im}^{\dagger}X_{\bm{r}i,m0}+X_{\bm{r}i,0m}\hat{b}_{\bm{r},im}\right)-X_{\bm{r}i,00}\sum_{m=1}^{d-1}\hat{b}^{\dagger}_{\bm{r},im}\hat{b}_{\bm{r},im}
+\sum_{m=1}^{d-1}\sum_{n=1}^{d-1}\hat{b}^{\dagger}_{\bm{r},im}X_{\bm{r}i,mn}\hat{b}_{\bm{r},in}$$
where $X_{\bm{r}i,mn}=\left<\psi_{\bm{r}i,m}\right|\hat{X}\left|\psi_{\bm{r}i,n}\right>$. The bosonic operators follow the commutation relations:$$
\left[\hat{b}_{\bm{r},im},\hat{b}_{\bm{r}',jn}\right]=\left[\hat{b}^{\dagger}_{\bm{r},im},\hat{b}^{\dagger}_{\bm{r}',jn}\right]=0$$
$$
\left[\hat{b}_{\bm{r},im},\hat{b}^{\dagger}_{\bm{r}',jn}\right]=\delta_{\bm{r}\bm{r}'}\delta_{ij}\delta_{mn}$$

Now for each interaction term in the Hamiltonian, we can expand the operators and only keep the quadratic terms. For example, considering the $x$ component of the $J_1$ Heisenberg interaction between atom $\left(\bm{r},1\right)$ and $(\bm{r}',4)$:

$$
\begin{aligned}
&\quad J_1S^x_{\bm{r},1} S^x_{\bm{r}',4} \\ &\approx J_1\left[S^x_{\bm{r}1,00}+\sum_{m=1}^{d-1}\left(\hat{b}_{\bm{r},1m}^{\dagger}S^x_{\bm{r}1,m0}+S^x_{\bm{r}i,0m}\hat{b}_{\bm{r},1m}\right)-S^x_{\bm{r}1,00}\sum_{m=1}^{d-1}\hat{b}^{\dagger}_{\bm{r},1m}\hat{b}_{\bm{r},1m}+\sum_{m=1}^{d-1}\sum_{n=1}^{d-1}\hat{b}^{\dagger}_{\bm{r},1m}S^x_{\bm{r}1,mm'}\hat{b}_{\bm{r},1n}\right]\\
&\quad \quad \left[S^x_{\bm{r}'4,00}+\sum_{n=1}^{d-1}\left(\hat{b}_{\bm{r}',4n}^{\dagger}S^x_{\bm{r}'4,n0}+S^x_{\bm{r}',0n}\hat{b}_{\bm{r}',4n}\right)-S^x_{\bm{r}'4,00}\sum_{n=1}^{d-1}\hat{b}^{\dagger}_{\bm{r}',4n}\hat{b}_{\bm{r}',4n}+\sum_{m'=1}^{d-1}\sum_{n'=1}^{d-1}\hat{b}^{\dagger}_{\bm{r}',4m'}S^x_{\bm{r}'4,m'n'}\hat{b}_{\bm{r}',4n'}\right]\\
\end{aligned}
$$
We only keep the quadratic terms of $\hat{b}$ and $\hat{b}^{\dagger}$, and get in total $2+2+2\times2=8$ terms:
$$
\begin{aligned}
&\quad J_1S^x_{\bm{r},1} S^x_{\bm{r}',4} \\ 
&\approx J_1  \left[ S^x_{\bm{r}1,00}\sum_{m'=1}^{d-1}\sum_{n'=1}^{d-1}\hat{b}^{\dagger}_{\bm{r}',4m'} S^x_{\bm{r}'4,m'n'}\hat{b}_{\bm{r}',4n'} + S^x_{\bm{r}'4,00}\sum_{m=1}^{d-1}\sum_{n=1}^{d-1}\hat{b}^{\dagger}_{\bm{r},1m}S^x_{\bm{r}1,mn}\hat{b}_{\bm{r},1n}\right.
\\
&\quad \quad -S^x_{\bm{r}1,00}S^x_{\bm{r}'4,00}\sum_{n=1}^{d-1}\hat{b}^{\dagger}_{\bm{r}',4n}\hat{b}_{\bm{r}',4n}-S^x_{\bm{r}'4,00 }S^x_{\bm{r}1,00}\sum_{m=1}^{d-1}\hat{b}^{\dagger}_{\bm{r},1m}\hat{b}_{\bm{r},1m} \\
&\quad \quad \left.+\sum_{m=1}^{d-1}\sum_{m'=1}^{d-1}\left(\hat{b}_{\bm{r},1m}^{\dagger}S^x_{\bm{r}1,m0}+S^x_{\bm{r}i,0m}\hat{b}_{\bm{r},1m}\right)\left(\hat{b}_{\bm{r}',4n}^{\dagger}S^x_{\bm{r}'4,n0}+S^x_{\bm{r}',0n}\hat{b}_{\bm{r}',4n}\right) \right]
\end{aligned}
$$

We can also do the same thing for the ring exchange and get $6+6+2\times2\times \dbinom{6}{2} =72$ terms.
Then we will Fourier transform and define 
$$ \begin{aligned}
\hat{b}_{\bm{k}im}=\frac{1}{\sqrt{N}}\sum_{\bm{r}}\hat{b}_{\bm{r},im}\e^{-\ii\bm{k}\cdot\bm{r}} \\
\hat{b}_{\bm{k}im}^{\dagger}=\frac{1}{\sqrt{N}}\sum_{\bm{r}}\hat{b}_{\bm{r},im}^{\dagger}\e^{\ii\bm{k}\cdot\bm{r}}
\end{aligned} $$
where $N$ is the normalization factor which gives the commutation relation $\left[\hat{b}_{\bm{k}im},\hat{b}_{\bm{k'}i'm'}^{\dagger}\right]=\delta_{\bm{k}\bm{k}'}\delta_{ii'}\delta_{mm'}$. After collecting only the quadratic terms and Fourier transformation, we can get the Hamiltonian in the form $\sum_{\bm{k}}\mathbf{X}^{\dagger}_{\bm{k}}\mathbf{H}_{\bm{k}}\mathbf{X}_{\bm{k}}$, where $\mathbf{X}_{\bm{k}}$ is a $2\times4\times(d-1)=88$ dimension column vector $\mathbf{X}_{\bm{k}}=\left(\hat{b}_{\bm{k}1,1},\cdots,\hat{b}_{\bm{k}1,11},\hat{b}_{\bm{k}2,1},\cdots,\hat{b}_{\bm{k}4,11},\hat{b}_{-\bm{k}1,1}^{\dagger},\cdots,\hat{b}_{-\bm{k}4,11}^{\dagger}\right)^{\mathsf{T}}$, and thus $\mathbf{H}_{\bm{k}}$ is a $88\times88$ Hermitian semi-positive-definite (semi- when and only when the infimum of the magnon spectrum is zero) matrix. To get the excitation energy, we will diagonalize the Hamiltonian by a method proposed by Colpa \cite{COLPA1978327}, which is also used in linear spin wave theory \cite{linearspinwavetheory}. 

\begin{figure}[t]
  \centering
  \includegraphics[width=0.8\textwidth]{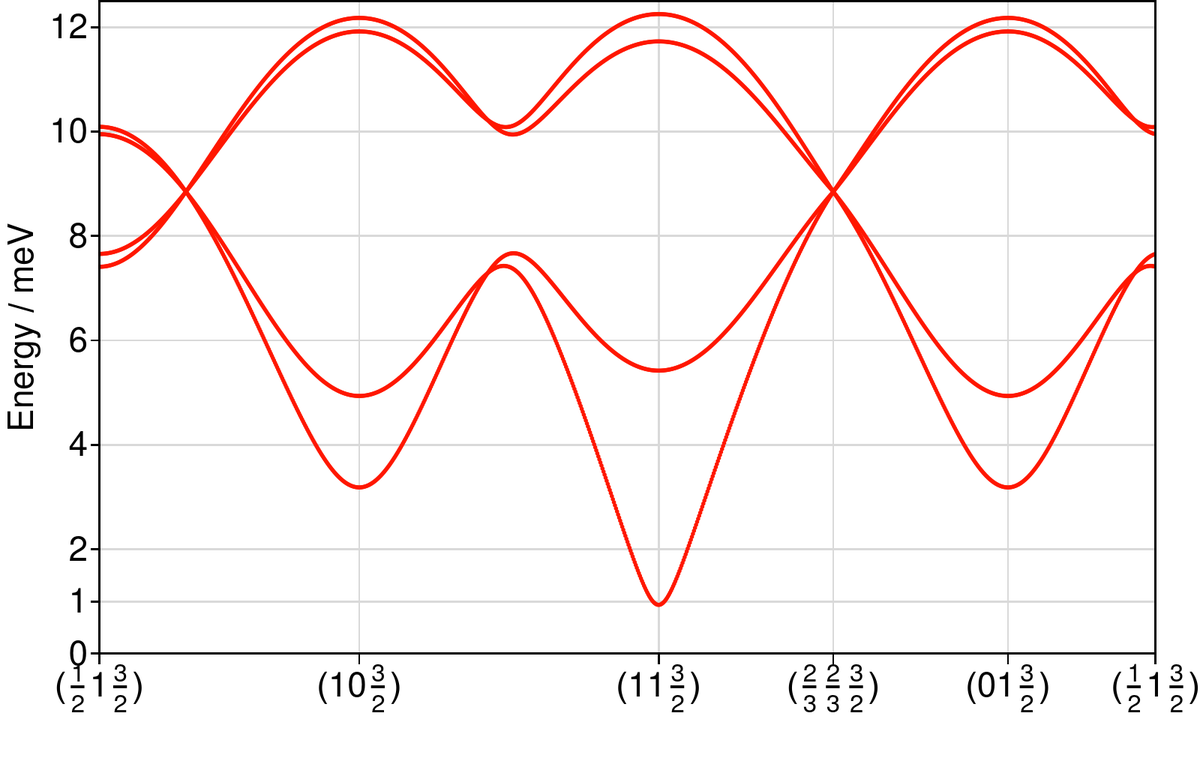}
  \caption{Magnon dispersion calculated from flavor wave model. This should be compared to Fig. 1 of \cite{neutron2021}}
  \label{fig:magnondispersion}
\end{figure}

We are able to fit the $\SI{0}{\tesla}$ to $\SI{22}{\tesla}$ Raman scattering data, data from ref. \cite{neutron2021}, and from refs. \cite{yuan2020dirac,yuan2020spin} with only six non-zero parameters. In all the calculations in both the main text and this supplementary note, they are (all in $\SI{}{\milli\electronvolt}$): trigonal distortion $\delta=52$, spin-orbital coupling $\lambda=16.4$, Heisenberg exchange $J_1=-0.90$, $J_4=0.189$, ring exchange $r=0.00062$ (with $\phi_6=\text{\textpi}$), and the biquadratic term $q=-0.15$. Because the calculated results with only $J_1$ and $J_4$ match the experiments well, the model with all $J_1$ through $J_6$ is not used. The calculated magnon dispersion is shown in figure \ref{fig:magnondispersion}.

\begin{figure}[htpb]
  \centering
  \includegraphics[width=0.8\textwidth]{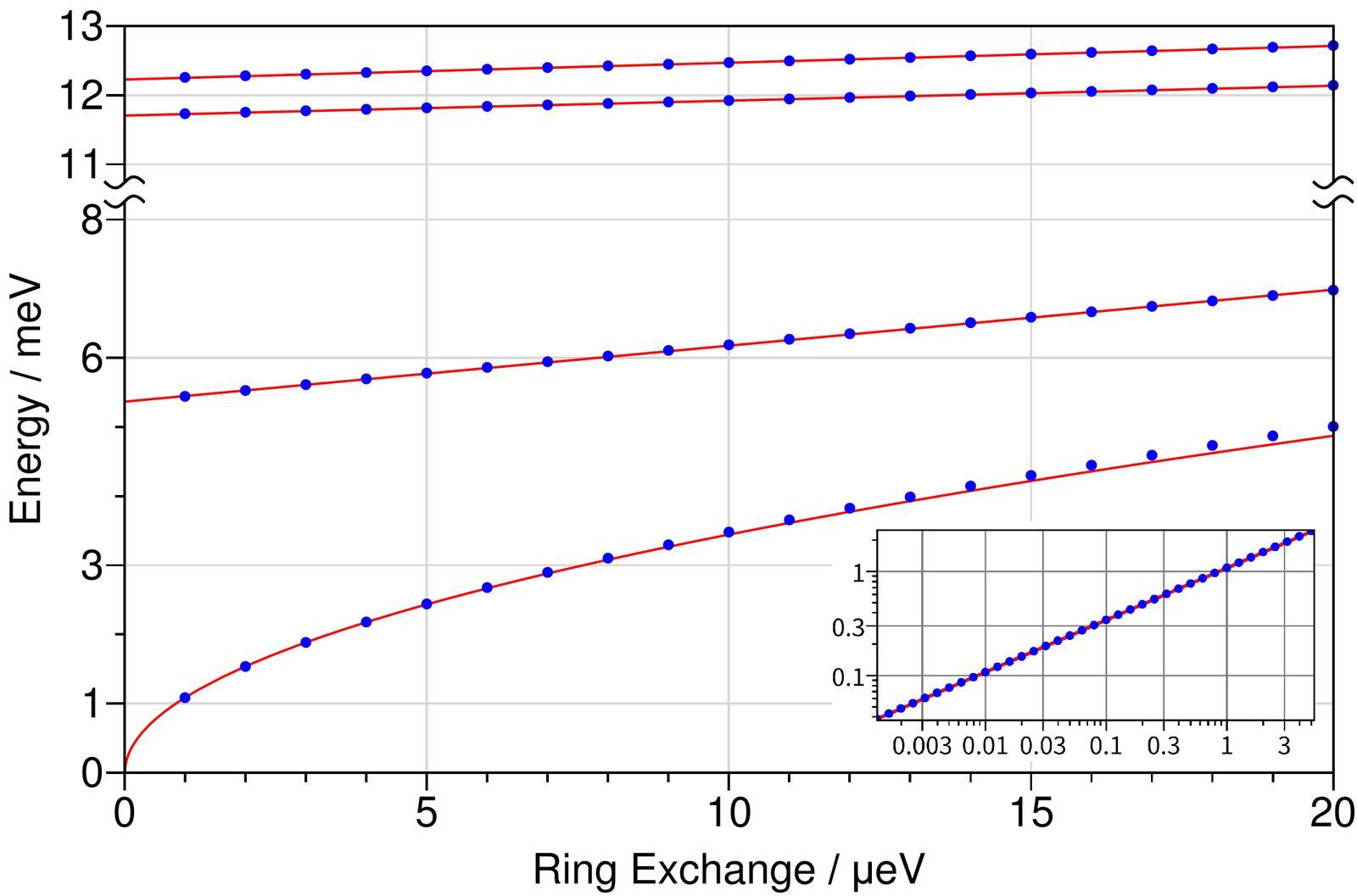}
  \caption{Magnon energy dependence on the ring-exchange interaction strength using the flavor wave model. Blue dots are calculated energy of the lowest 4 modes at the zone center. The lowest mode is fitted (solid line) with $E=\sqrt{Kr}$, $K=\SI{1.17}{\electronvolt}$. The other 3 modes are fitted with linear functions. Inset shows the behavior at a small ring exchange.}
  \label{fig:ringexchangedependence}
\end{figure}

\begin{table}
\centering
\caption{Gap opening multi-spin interactions. $E$ is the energy of the lowest magnon mode at the zone center. The dependence is fitted when $E<\SI{3}{\milli\electronvolt}$. Arg $r$ and Arg $p$ don't change the gap and can only determine where the spins are pointing in the $xOy$ plane. Even though the measurement gives $\SI{0.82(5)}{\milli\electronvolt}$ at zero field, the best fit gives $\SI{0.93}{\milli\electronvolt}$, and this gap is used in this table.}
\begin{tabular}[t]{c |c|c}
\hline
 Interaction & Dependence of the lowest mode &To get $\SI{0.93}{\milli\electronvolt}$ gap \\
 \hline
 \begin{minipage}{.35\textwidth}
 \includegraphics[width=1\textwidth]{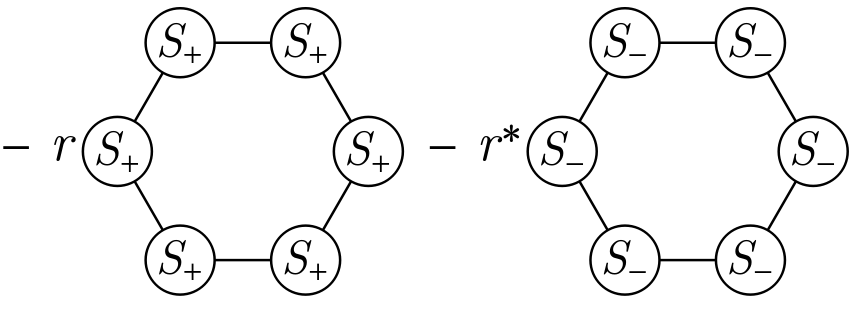} 
 \end{minipage}& $E=\sqrt{
 \SI{1.17}{\electronvolt}
 \times |r|}$&$r=\SI{0.74}{\micro\electronvolt}$  \\  
 \hline
 \begin{minipage}{.35\textwidth}
 \includegraphics[width=1\textwidth]{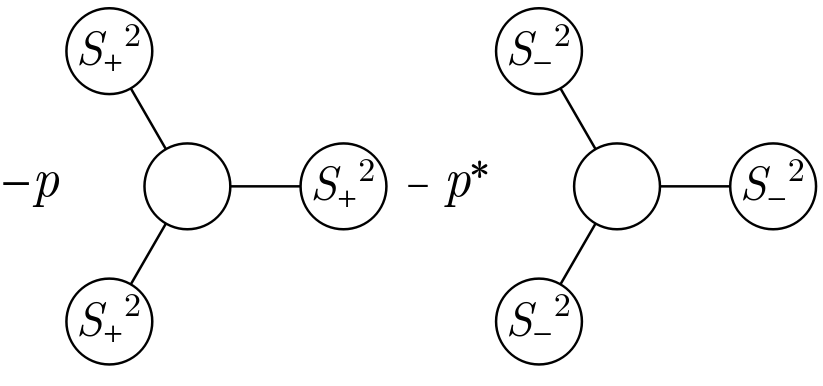} 
 \end{minipage}& $E=\sqrt{
 \SI{0.110}{\electronvolt}
 \times |p|}$&$|p|=\SI{7.9}{\micro\electronvolt}$  \\

 \hline
 \begin{minipage}{.35\textwidth}
 \includegraphics[width=1\textwidth]{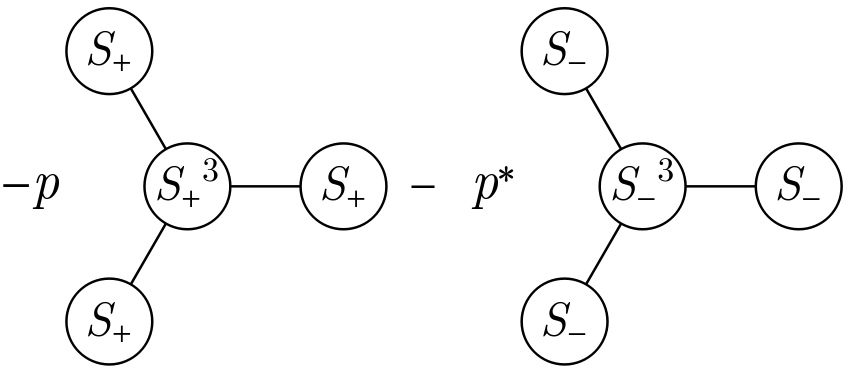} 
 \end{minipage}& $E=\sqrt{
 \SI{0.105}{\electronvolt}
 \times |p|}$&$|p|=\SI{8.3}{\micro\electronvolt}$  \\    
 \hline

\begin{minipage}{.35\textwidth}
\includegraphics[width=1\textwidth]{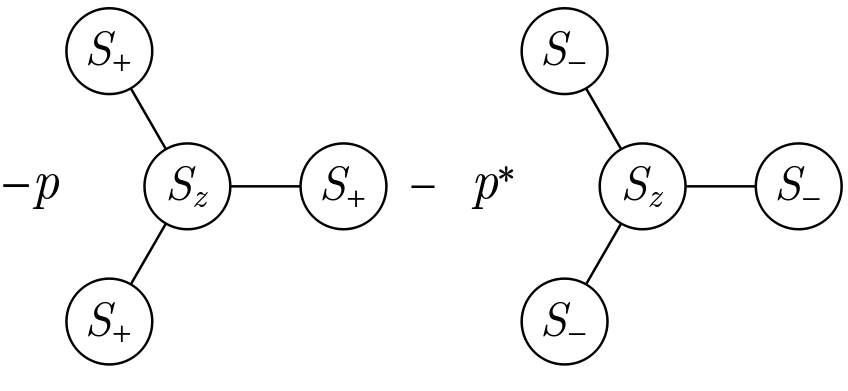} 
\end{minipage}& $E=\SI{13.4}{\milli\electronvolt}\times |p|$&$|p|=\SI{69}{\micro\electronvolt}$  \\   
 \hline
Nearest neighbour $i$,$j$: $p(S_i^+)^3(S_j^+)^3+p^*(S_i^-)^3(S_j^-)^3$& $E=\sqrt{
 \SI{7.0}{\milli\electronvolt}
 \times |p|}$& $|p|=\SI{0.12}{\milli\electronvolt}$  \\    
 \hline
\end{tabular}\label{tablesupl}
\end{table}

We note, however, that there are other possible interactions which are able to open the gap without breaking the $C_3$ symmetry. Some of these interactions are shown in table \ref{tablesupl}. Notice that to open the same gap magnitude, the interaction of the ring exchange is at least one order of magnitude smaller than the other interactions. Also, when the operators $(S^+)^2$, $(S^-)^2$, $(S^+)^3$ and $(S^-)^3$ are projected into the $\tilde{S}=1/2$ subspace, they all vanish. This means that the dominant contribution should only come from the ring exchange term, which still exists in the $\tilde{S}=1/2$ subspace. The interaction in the fourth row does not vanish in the $\tilde{S}=1/2$ subspace, but this will make the spins have a $z$ component($\pm 0.07_{\mu\mathrm{B}}$) in the ground state but this has not been reported from neutron scattering experiments.

In figure \ref{fig:ringexchangedependence} we show the dependence of the magnon energies at zero magnetic field and at the Brillouin zone center for selected values of the ring exchange parameter $r$ (dots). Together with the calculations we also show fits to this dependence as solid lines. The gap of the pseudo-Goldstone mode is best fit as $E_{\mathrm{gap}}\propto\sqrt{r}$ when $r$ is small. Note that this dependence is the same as found in the $\tilde S = 1/2$ model, as shown in figure \ref{fig:ringdep}. 
\begin{figure}[t]
  \centering
  \includegraphics[width=0.8\textwidth]{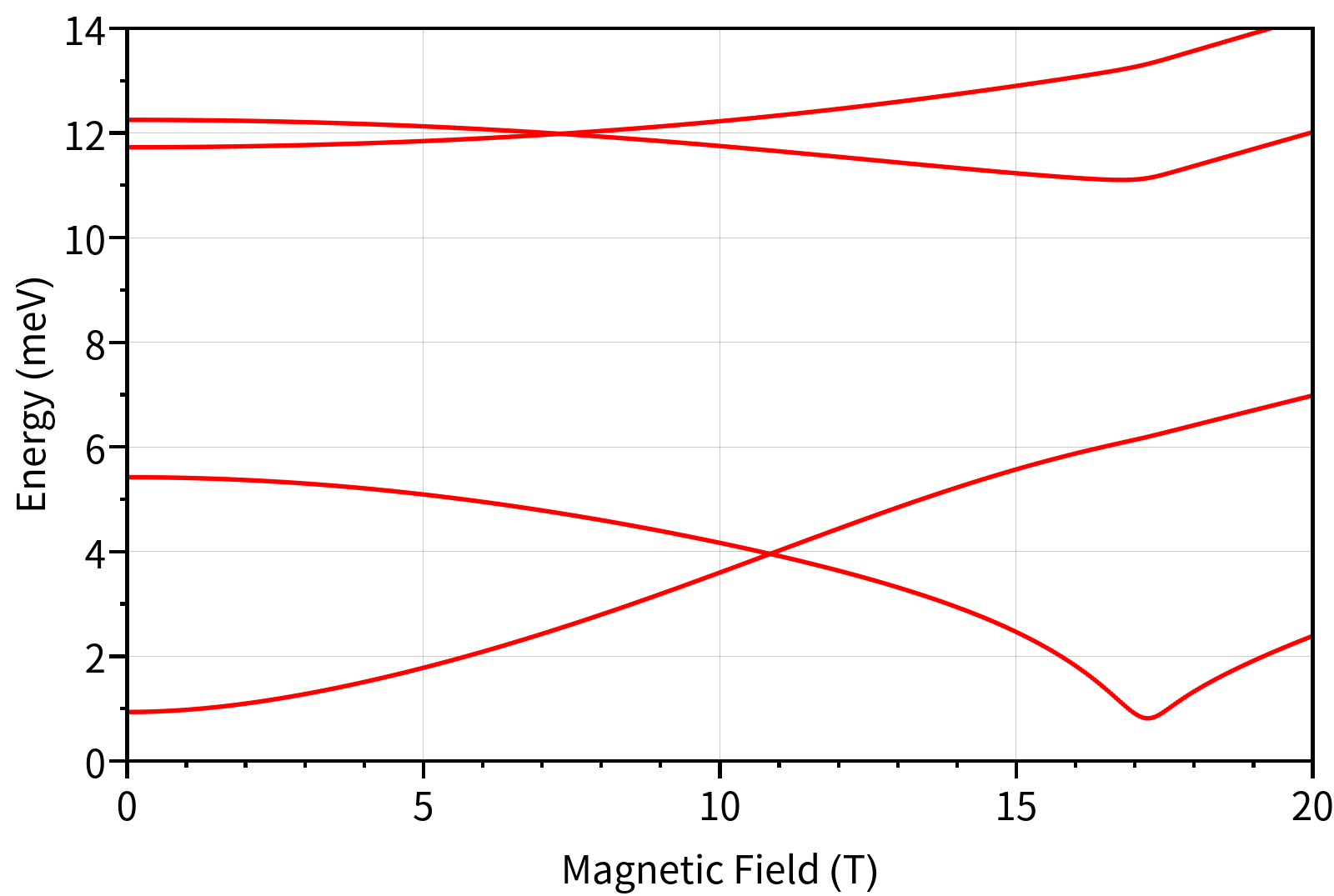}
  \caption{Zone center magnon energy dependence on magnetic field in the flavor wave model. Magnetic field is assumed to be perpendicular to the spins at zero field.}
  \label{fig:magneticfielddependence}
\end{figure}

Finally, we calculated the magnetic field dependence of the four magnons at the zone center for magnetic fields parallel to the honeycomb plane, shown in figure \ref{fig:magneticfielddependence}. We can compare these with the calculation in figure \ref{fig:s1/2fielddep}, which shows agreement on the field dependence of the lowest two modes below $\SI{9}{\tesla}$. As shown in figure 3(\textbf{B}) of the main manuscript, both models continue to agree similarly well with the behavior at higher fields.

\bibliography{CTO_Biblio}% Produces the bibliography via BibTeX.